\documentclass[letterpaper, 12pt]{amsart}
\usepackage{amsthm}
\usepackage{amssymb}
\usepackage{caption}
\usepackage[labelformat=simple,labelfont={}]{subcaption}

\usepackage[utf8]{inputenc}
\usepackage[dvips]{graphicx}
\usepackage{epstopdf}
\usepackage{natbib}
\usepackage{soul,color}
\usepackage{hyperref}
\usepackage{comment}
\hypersetup{
     colorlinks = true,
     linkcolor = blue,
     anchorcolor = black,
     citecolor = black,
     filecolor = blue,
     urlcolor = blue,
     breaklinks=true}

\def\RR{\mathbb{R}}    
\def\PP{\mathbb{P}}    
\def\EE{\mathbb{E}}    
\def\R{{\mathcal R}}   

\def\proof{\noindent{\em Proof.}~}
\def\eproof{\mbox{\ }\hfill$\square$}
\def\tcb{\textcolor{blue}}
\def\tcr{\textcolor{red}}

\newtheorem{lemma}{Lemma}
\newtheorem{proposition}{Proposition}

\newtheorem{definition}{Definition}
\newtheorem{assumption}{Assumption}

\title{A recursive logit model with choice aversion and its application to transportation networks \\ 
\tiny{({\normalfont Forthcoming at Transportation Research Part B: Methodological})}}
\author{Austin Knies  \quad Jorge Lorca\quad Emerson Melo}
\thanks{ The results in this paper were circulated earlier as two separate papers: “Choice aversion in directed networks,” by Lorca and Melo, and “A recursive logit model with choice aversion and its application to path choice analysis,” by Knies and Melo. Please address correspondence to: Emerson Melo, Department of Economics, Indiana University, 307 Wylie Hall, 100 S Woodlawn Ave Bloomington, IN 47408, U.S.A. Email: emelo@iu.edu. Austin Knies,  Department of Economics,  Indiana University-Bloomington, ausknies@iu.edu. Jorge Lorca, Central Bank of Chile. The views expressed herein are those of the authors and do not necessarily represent the opinion of Central Bank of Chile or its board. We are very grateful to Mogens Fosgerau, Emma Frejinger, and  Tien Mai for facilitating the codes used in this paper. This research was partially funded by the National Science Foundation NRT grant 1735095, ``Interdisciplinary Training in Complex Networks and Systems." We are also very grateful to the Associate Editor and three anonymous referees for their very valuable comments and suggestions that greatly improved the paper.}
\date{\today}

\AtBeginDocument{}%
 
\begin{document}

\maketitle

\pagestyle{myheadings} \thispagestyle{plain} \markboth{ }{ }
\begin{abstract}   
	We propose a recursive logit model which captures the notion of choice aversion by imposing a penalty term that accounts for the dimension of the choice set at each node of the transportation network. We make three contributions. First, we show that our model overcomes the correlation problem between routes, a common pitfall of traditional logit models, and that the choice aversion model can be seen as an alternative to these models.
	 Second, we show how  our model can generate violations of regularity in the path choice probabilities. In particular, we show that removing edges in the network may decrease the probability for existing paths.
	 Finally, we show that under the presence of choice aversion, adding edges to the network can make users worse off. In other words, a type of Braess's paradox can emerge outside of congestion and can be characterized in terms of a parameter that measures users' degree of choice aversion. We validate these contributions by estimating this parameter over GPS traffic data captured on a real-world transportation network.
\end{abstract}


\thispagestyle{empty}


\section{Introduction}\label{introduction}

Discrete choice models have been used extensively to understand the behavior of  participants (who we will refer to as users) in transportation networks (\citet{mcffadens,mcf1} and \citet{BenAkiva_Lerman1985}). In this context, users  choose a path  that maximizes total utility (or, equivalently, minimizes their total cost) of traveling. A prominent model that has arisen from this literature is the Multinomial Logit (MNL) Model,  whose main advantage  is its tractability and closed-form choice probabilities.
\smallskip

Despite its popularity, the MNL presents some drawbacks when applied to the case of transportation networks. In particular, the MNL can predict unrealistic choice probabilities for paths sharing common edges in the network. The root of this problem traces back to the  Independence of Irrelevant Alternatives (IIA) axiom that is  required to derive the MNL. 
\smallskip

\begin{figure}[h]
	\centering
	\includegraphics[width=0.49\textwidth]{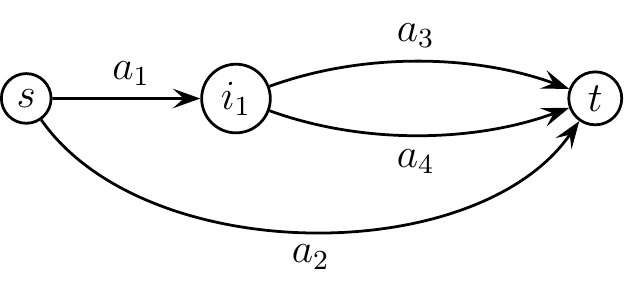}
	\caption{Paths $(a_1,a_3)$ and $(a_1,a_4)$ share edge $a_1$.}
	\label{fig:Logit_Path_choice_intro}
\end{figure}

To explain the severity of  the overlapping paths problem, consider  the simple transportation network displayed in  Figure \ref{fig:Logit_Path_choice_intro}.\footnote{This example follows the discussion in \citet{BenAkivaBierlarie1999}.}

\smallskip
In this network we have three paths: $(a_1,a_3)$, $(a_1,a_4)$, and $(a_2)$. Assume the utility of each path is 1. At the edge level, assume that the utility of edge $a_1$ is $1-\theta$, the utility of edges $a_3$ and $a_4$ is $\theta$, and, finally, the utility of edge $a_2$ is 1. The value of $\theta$ is assumed to satisfy $0<\theta<1$.
\smallskip

For the setting described above, the MNL will predict that each path is chosen with probability $1/3$. However, these choice probabilities are unrealistic when paths lack distinctiveness or independence from another. In particular,  an assignment in which path $(a_2)$ is chosen with probability $1/2$ and  paths $(a_1,a_3)$ and $(a_1,a_4)$ are chosen with probability 1/4 is more sensible. More explicitly, this latter solution takes into account the fact that paths $(a_1,a_3)$ and $(a_1,a_4)$ share the common edge $a_1$ and in terms of total cost they are equivalent to each other as well as the cost of path $(a_2)$. The reason why the MNL model cannot accommodate situations where  the  path costs  are correlated is that the MNL relies on the property of Independence of Irrelevant Alternatives (IIA).\footnote{We  remark  that whether the IIA property holds or not in the network discussed above does not depend on the choice of utilities associated to each edge. In fact, a similar conclusion holds when we allow for general link utilities. The violation of IIA is a result that paths $(a_1,a_3)$ and $(a_1,a_4)$ share the common link $a_1$. This latter fact is what generates the failure of IIA. } As the discussion above shows, the IIA property is hardly satisfied even in simple transportation networks.\footnote{An in-depth discussion on path correlation and IIA can be found in \citet{mcfadden1974frontiers}, which discusses the canonical red bus/blue bus problem.} 

\smallskip

Recognizing this pitfall, the transportation literature has proposed several \emph{corrections} to the MNL. This   class of extensions is known as Path Size Logit (PSL). The idea of this class of models  is to correct the problem of overlapping paths by adding an extra correlation-penalizing term to path costs. Thus, when the choice probabilities are generated through a standard MNL, the correction will account for the degree of overlapping between different paths.
\smallskip

While its usefulness in correction is clear, the PSL class has two problems. First, the type of correction employed by the different models do not have a theoretical justification in terms of users' behavior. In particular, the parameters describing the corrections do not have a direct interpretation from an economic viewpoint. Second, it is not clear how PSL models can be used to carry out welfare analysis in transportation networks. For instance, it is hard to  interpret  and predict the changes on welfare when edges are added to, or severed from, the network.

In this paper, we propose the use of a recursive logit (RL) model which incorporates the idea of choice aversion  (choice overload) in users' behavior. In doing so we adapt  the approach of  \citet{FudenbergStra2015} to the context of directed graphs. Simply put, the choice aversion hypothesis states that an increase in the number of alternatives to choose from may lead to adverse consequences, such as lesser motivation to actually choose or lower satisfaction ex post (cf. \citet{SheenaLepper2000} and \citet{Scheibehenneetal2010}).

Formally, we consider a transportation network  with \emph{source} node  $s$ and designated \emph{sink} node $t$. In this setting, we model users' behavior as a sequential choice process: when assessing an edge $a$ at some node $i\neq t$, users evaluate both the \emph{flow} utility and the appropriate \emph{continuation} value associated to such an edge. Following \citeauthor{FudenbergStra2015},  we introduce a term that penalizes the size of each choice set that stems subsequently from every current edge under scrutiny.  In particular, when considering an edge $a$ at node $i$, users will penalize the number of outgoing edges at $i$. In other words, when facing a set of alternatives in order to depart from a specific node, users incorporate the size of the ensuing choice set when they appraise the continuation value of each outgoing edge. Formally, at each node $i\neq t$, we consider the penalty $\kappa_i\log|A_i^+|$, where $|A_i^+|$ is the cardinality of the set of outgoing edges at node $i$ and $\kappa_i\geq 0$ is a parameter that captures users' choice aversion degree at node $i$.
\smallskip

We make three contributions. First, we show that RL with choice aversion  can overcome the problem of overlapping in transportation networks. In particular, the parameters $\kappa_i$ plays a critical role in the  form of the correction. We show that our model performs as well as the recent Adaptive PSL model introduced by \citet{DUNCAN20201}. However, our correction has two main advantages. First, it is a simple correction based on users' optimal behavior. Second, the parameters $\kappa_i$ have a clear interpretation in terms of users' attitude with respect to  size of choice sets.
\smallskip

In our second contribution, we show how our model  captures  violations of regularity (\citet{LuceSuppes1965}). Formally, we show  that removing an edge in a particular node can decrease the choice probabilities of some  paths in the network. To grasp how this result works,  we note that  removing an edge $a$ at node $i$ is equivalent to removing the set of all  paths in which  $a$ is a member. When removing an edge in the traditional MNL (i.e.,  the choice aversion model with $\kappa_i=0$ for all $i\neq s,t$),  the choice probability of the remaining paths increase proportionally.
	 However, in our model,  removing the edge $a$ not only reduces the set of available paths (passing through node $i$) but also decreases the choice aversion costs associated to these paths. This latter reduction makes the set of paths passing through node $i$ comparatively more valuable than those not using node $i$. As a consequence, the path choice  probabilities  of  paths  passing through node $i$ may increase due to  the reduction in  the  set of available paths  and  the  choice aversion cost reduction.
 \smallskip
 
  We formalize  the failure of regularity  in terms of a  precise relationship between the parameters $\kappa_i$  and the choice probabilities of paths  using the node where the edge is removed.  As far as we know, this result is new to the literature on recursive models in transportation networks. From a practical standpoint, accounting for the failure of regularity provides useful information to planners who must predict flows in transportation networks.

\smallskip

In our final contribution, we study  how our model can capture a Braess's-like paradox. In particular, we show how adding edges to the network can decrease users' welfare. Similarly to regularity failure, we also characterize this result in terms of  the degree of choice aversion $\kappa$. Unlike existing models and extensions, the behavioral foundation of the choice aversion model and its formulation allows for a tradeoff between instantaneous route utilities and choice aversion penalization such that decreases in users' welfare can be observed even in the absence of congestion. To our knowledge, this is also a novel result that can shed light on the design of  transportation networks and its effects on welfare.

To ground these contributions, we conclude with an empirical analysis of the choice aversion model and measure its performance relative to other RL models. The results of these maximum likelihood estimation routines provide empirical support that real-world network users experience choice overload and display choice averse behavior. Additionally, we show that the simplicity of the choice aversion model allows for the model to be estimated much more quickly than other RL models, all while providing similar  corrections to path predictions as well as maintaining a clear economic and behavioral interpretation.
\subsection{Related Literature} RL models have  been studied by \citet{bc}, \citet{Fosgerauetal2013}, \citet{MAI2015100}, and \citet{Mai2016}.\footnote{For an up to date survey on recursive models in traffic networks we refer the reader to \citet{ZIMMERMANN2020100004}.}  Our paper differs from theirs in at least two dimensions. First, we extend their recursive approach to incorporate the notion of choice aversion. Second, we show how our model can handle the problem of path overlapping, violations of regularity, and Braess's-like paradoxes. It is worth pointing out that the papers by \citet{MAI2015100} and \citet{Mai2016} correct the overlapping problem by using a Nested Recursive Logit (NRL)  and  a Generalized Recursive Extreme Value model, which allow for correlation among the stochastic terms. Our approach preserves the independence of the random terms, where the choice aversion penalty allows us to capture and correct the  degree of overlapping between paths. 

 With respect to PSL models, the existing literature is extensive (\citet{BenAkivaBierlarie1999} and \citet{Frejinger_Bierlarie2003}), and  \citet{DUNCAN20201} present an up-to-date discussion of the PSL approach.\footnote{We must  also mention that the papers by \citet{Chu1989}, \citet{KoppelmanChieh2000}, \citet{Vosha1997}, \citet{Vosha1998}, and \citet{WenKoppelman2001} propose  variants of the MNL model in an attempt to generate flexible discrete choice models that allow for correlation between different paths.} In addition, they propose an alternative correction denominated as the \emph{Adaptive Path Size Logit} (APSL) model. Our results differ from theirs in the type of the behavioral foundation we use.
\smallskip

 From a behavioral standpoint,  a similar approach to this paper is found in \citet{FOSGERAU20191} and \citet{JIANG2020188}, who incorporate a rational inattention model into the context of transportation networks. While choice aversion and rational inattention are interconnected in terms of information processing, we model a particular variant of costly decision-making in the form of aversion to increasing choices, rather than mutual information through observing a signal. This modeling choice allows us to generate clear path choice predictions, violations of regularity, and Braess's-like paradox phenomena.

Finally, we mention that our paper is related to \citet{Acemogluetal2017}. They study a congestion game introducing the notion of information-constrained Wardrop Equilibrium, which allows users to possess different information sets about the congestion level in the network. They show that providing more information to users can make them worse off.  Our work differs from theirs in at least three aspects. First, we focus on the notion of choice aversion while they focus on the concept of information sets. Second, we show our Braess's-like result in the context of a RL model, while they study a deterministic and non-recursive model. Finally, we show that Braess's paradox may occur even in uncongested networks while they focus in the case of congested networks.

The rest of the paper is organized as follows. Section \ref{s2} introduces the RL model with choice aversion. Section \ref{s3} explores the use of choice aversion in the path choice model and in comparison to existing Path Size Logit (PSL) models. Section \ref{s4}  discusses the failure of regularity. Section \ref{s5} discusses a type of Braess's paradox observed as a consequence of choice aversion. Section \ref{s6} conducts an empirical analysis to estimate the role of choice aversion in real-world contexts. Section \ref{s7} concludes. Technical details, proofs, and further welfare analyses are gathered in Appendixes \ref{Appendix}, \ref{appendix:proofs}, and \ref{appendix:welfarex}, respectively.

\section{Recursive logit  in Directed Networks}\label{s2}

In this section we propose a  recursive discrete choice model in directed networks. Formally, we model a set of users as solving a dynamic programming problem over a directed, acyclic graph. In a noticeable departure from previous literature, we adapt the choice aversion formulation of \citet{FudenbergStra2015} into the context of directed graphs, and then we analyze the consequences on equilibria and welfare.\footnote{Throughout the text we use the term choice overload to refer to choice aversion.}
\subsection{Directed graphs} Consider a directed  (not assumed acyclic) graph $G = (N,A)$ where $N$ is the set of nodes  and $A$ the set of edges, respectively.  We denote the set of \emph{ingoing} edges to node $i$ by $A_i^-$, and the set of \emph{outgoing} edges from node $i$ by $A_i^+$. We refer accordingly to the \emph{out-degree} of node  $i$ as $|A_i^+|$.

Without loss of generality, we assume that $G$ has a single source-sink pair, where $s$ and $t$ stand for the \emph{source} (origin) and \emph{sink} (destination) nodes, respectively. Let $j_a$ be the node $j$ that has been reached through edge $a$. We therefore define a path as a  sequence of edges $(a_1,\ldots a_K)$ with $a_{k+1} \in A_{j_{a_{k}}}^+$ for all $k< K$. 

The  set of paths  connecting nodes $s$ and  $t$ is denoted by  $\mathcal{R}$. The  set of paths connecting  nodes $s$ and $i\neq t$  is denoted by $\mathcal{R}_{si}$. Similarly, the set of all paths connecting nodes $i\neq s$ and $t$ is denoted as $\mathcal{R}_{it}$.  Let $\R_i$ denote the set of paths passing through $i$. Finally,  let $\R_i^c$ denote  the set of paths not passing through node $i$.\footnote{We note that $\R_i, \R_{si},\R_{it}$, and $\R_i^c$ are subsets of $\R$. }
\smallskip

 A deterministic utility component $u_a>0$ is associated with each edge $a\in A_i^+$ for all $i\neq t$.  Path utilities are assumed to be edge additive; that is, for a path $r=(a_1,\ldots a_K)\in \R$, its associated utilities are given by  $\sum_{k=1}^Ku_{a_k}$.

We assume that at  node $s$ there is a unitary mass of network users who must choose a path from the set $\R$. For the sake of exposition,  the mass of users is summarized by the canonical vector $e_s$,  which has a 1 in the position of node $s$ and zero elsewhere. The dimension of  $e_s$  is  $|N|-1$.
\subsection{Choice aversion}  We now develop a RL choice model  over $G$ that incorporates choice overload by means of an specific kind of penalty on ensuing choice sets stemming from each edge appraisal. In particular, we adapt the choice aversion approach from \citet{FudenbergStra2015}  into the environment described by $G$ as follows: for each $a\in A^+_i$ we associate  a collection of i.i.d.  random variables $\{\epsilon_{a}\}_{a\in A^+_{i}}$ such that the \emph{recursive} utility associated to edge $a$  is defined as:
\begin{equation}\label{recursive_utility}
V_a=u_a+\EE\left(\max_{a^\prime \in  A^+_{j_a}}\{V_{a^\prime}+\epsilon_{a^\prime}-\kappa_{j_a}\log |A^+_{j_a}| \}\right)\quad \mbox{ for all $a\in A^+_{i}$},
\end{equation}
where  $u_a$ denotes the \emph{instantaneous} utility associated to edge $a$ and  the term  $$\EE\left(\max_{a^\prime \in  A^+_{j_a}}\{V_{a^\prime}+\epsilon_{a^\prime}-\kappa_{j_a}\log |A^+_{j_a}| \}\right)=\EE\left(\max_{a^\prime \in  A^+_{j_a}}\{V_{a^\prime}+\epsilon_{a^\prime}\}\right)-\kappa_{j_a}\log |A^+_{j_a}| $$ is the \emph{adjusted} continuation value associated to the selection of $a$. Notice that the latter term includes the factor $\kappa_{j_a}\log |A^+_{j_a}|$, which is a penalty term that captures the size of the set $A^+_{j_a}$, where $\kappa_{j_a}\geq 0$.\footnote{\citet{FudenbergStra2015}  study a RL model in the context of intertemporal choice. In doing so, they consider a discount factor $\delta\in (0,1)$. We focus on a digraph $G$ without discounting.}
\smallskip

Following \citet{FudenbergStra2015}, we impose the following assumption on the random variables $\{\epsilon_{a}\}_{a\in A^+_{i}}$.
\begin{assumption}[Logit choice rule]\label{Gumbel}  At each node $i\neq t$ the collection of random variables $\{\epsilon_a\}_{a\in A_{i}^+}$ follows a Gumbel distribution with scale  parameter $\mu=1$. 
	\end{assumption}

Under this  assumption,  Eq. (\ref{recursive_utility}) can be expressed as:
\begin{equation}\label{recursive_utility2}
V_a=u_a+\mbox{$\log\left(\sum_{a^\prime \in  A^+_{j_a}}e^{V_{a^\prime}}\right)$}-\kappa_{j_a}\log |A^+_{j_a}|,
\end{equation}
where  $\log\left(\sum_{a^\prime \in  A^+_{j_a}}e^{V_{a^\prime}}\right)-\kappa_{j_a}\log |A^+_{j_a}|$ provides a closed-form expression for the adjusted continuation value.\footnote{See \citet{Train1}, Chapter 3.}

 Let us define $\varphi_{j_a}(V)\triangleq \log\left(\sum_{a^\prime \in  A^+_{j_a}}e^{V_{a^\prime}}\right)$   and $\hat{\varphi}_{j_a}(V)\triangleq\varphi_{j_a}(V)-\kappa_{j_a}\log|A_{j_a}^+|$ for all  $j_a\neq t$. Accordingly, Eq. (\ref{recursive_utility2}) can be rewritten as:
\begin{equation}\label{recursive_utility3}
V_a=u_a+\hat{\varphi}_{j_a}(V).
\end{equation}

The previous expression deserves some remarks. First, the adjusted continuation value in Eq. (\ref{recursive_utility3}) captures the  \emph{complexity} of the choice sets   $A_{j_a}^+$,  as measured by  $\kappa_{j_a}\log |A^+_{j_a}|$,  with $\kappa_{j_a}\geq 0$. Intuitively,  the expression $\kappa_{j_a}\log |A^+_{j_a}|$ penalizes the size of the choice sets at different nodes, where the  parameter $\kappa_{j_a}\geq 0$  measures users' attitudes towards the size of $A_{j_a}^+$. In particular, $V_a$ is a decreasing function of $\kappa_{j_a}$.

Second, when $\kappa_{j_a}=0$ for $j_a \notin \{s,t\}$,  the expression in Eq. (\ref{recursive_utility3}) boils down to a traditional RL model where users are \emph{choice-loving}: they \emph{always} prefer to add additional items into the menu, as in the “\emph{preference for flexibility}” of \citet{Kreps1979}.  To see this, note when $\kappa_{j_a}=0$,  the function $\varphi_{j_a}(V)$ is increasing in $|A_{j_a}^+|$. As a consequence,  the recursive utility $V_a$  increases  as the size of  $A_{j_a}^+$ increases. This latter feature implies that traditional RL models in transportation networks (e.g., \cite{bc}, \cite{Fosgerauetal2013}, \citet{MAI2015100}, and \citet{Mai2016}) can be associated with an intrinsic taste for plentiful options.

 On the other hand,  the case of  $\kappa_{j_a}\in(0,1)$  may be interpreted as a situation where the users prefer to include additional edges (alternatives) to the menu, provided the new options are not \emph{too much worse than the current average.}  The case of $\kappa_{j_a}=1$ captures  a situation where the users want to remove choices (links)  that are \emph{worse} than the average: they worry about choosing such additional alternatives by accident given \emph{appraisal} costs---such as \citet{Ortoleva13}'s thinking aversion---that may offset the benefits of the corresponding random draw associated to $\epsilon$.  Finally, the case of  $\kappa_{j_a}> 1$ should be interpreted as a situation in which the users only wish to add alternatives that are perceived to be sufficiently better.
 \smallskip
 
We remark that  the three cases of possible values that  $\kappa_{j_a}$ may take provide us the flexibility to capture different users' attitudes with respect to the size of the choice set at each node.  In sum, the collection of parameters $\{\kappa_{j_a}\}$ encapsulates the scale of penalties on the set of ensuing actions arising from each nonterminal node, which unlocks keen consequences on users' attitudes towards marginally increasing the set of edges.\footnote{In our analysis we focus on a digraph with a single origin-destination pair. However, our analysis can be adapted to the case where we have multiple origin-destination pairs. In particular, we can specify a parameter $\kappa_{j_a}^{s_l t_l}$ where $s_l t_l$ denotes a particular origin-destination pair for $l=1,\ldots, L.$}
 \smallskip

  Following \citet{FudenbergStra2015}, we identify $\{\kappa_{j_a}\}_{a\in A^+_{i}}$ as the users' choice aversion parameters. 
 
\subsection{Flow allocation} Each user is looking for an optimal path connecting $s$ and $t$. Now, when they reach node $i\neq t$, they observe the realization of the random utilities  $V_a+\epsilon_{a}$  for all $a\in A_i^+$, and consequently choose the  alternative $a\in A^+_{i}$  with the  highest utility. 
\smallskip

This process is repeated at each subsequent node giving rise to a recursive discrete choice model, where the expected flow  entering node $i\neq t$ splits among the alternatives $a\in A_i^+$ according to the choice probability:
\begin{equation}\label{choice}
\PP(a|A^+_i)=\PP\left(a=\arg\max_{a^\prime\in A_{i}^+}\{V_{a^\prime}+\epsilon_{a^\prime}\}\right)\quad \forall i\neq t.
\end{equation}

Due to Assumption \ref{Gumbel}, Eq. (\ref{choice}) can be rewritten as:
\begin{equation}\label{Logit_choice}
\PP(a|A^+_i)={{e^{u_a+\varphi_{j_a}(V)-\kappa_{j_a}\log |A_{j_a}^+|}}\over\sum_{a^\prime \in A_i^+} e^{u_{a^\prime}+\varphi_{j_{a^\prime}}(V)-\kappa_{j_{a^\prime}}\log |A_{j_{a^\prime}}^+|}}\quad \forall i\neq t.
\end{equation}

As $\kappa_{j_a}$ increases, the edge choice probability $\PP(a|A^+_i)$ is increasingly penalized by the size  of the choice set $|A_{j_a}^+|$, reflecting the cost of choice overload onto a user's edge utility from nodes with large choice sets.
This is  a fundamental difference with the traditional RL model, which assumes $\kappa_{j_a}=0$ as we mentioned before.

 Mathematically, the recursive process just described induces a Markov chain over  the graph $G$, where the transition probabilities are given by Eq. (\ref{Logit_choice}).
Let $x_i$ be the  expected flow entering at node $i$ towards sink node $t$. Then the flow received by edge $a$ is given by:
\begin{equation}\label{expected_flow}
f_{a}=x_i \PP(a|A^+_i)\quad\forall a\in A^+_{i},
\end{equation}
with $f=(f_a)_{a\in A}$ denoting the expected flow vector.
\smallskip

 In addition, let $\hat{\PP}=(\PP_{ij})_{i,j\neq t}$ denote the restriction to the set of nodes $N\setminus\{t\}$. Then the expected demand vector $x=(x_i)_{i\neq t}$ may be expressed 
as $x=e_s+\hat{\PP}^Tx$ which generates the following stochastic conservation flow equations
\begin{equation}\label{flow_conservation1}
x_i=\sum_{a\in  A_i^-}f_{a}\quad\mbox{for all $i\neq t$}.
\end{equation}
A flow vector $f$ satisfying (\ref{flow_conservation1}) is called \emph{feasible}. It is worth  mentioning that there exists a unique flow vector $x^*$ satisfying the flow constraints (\ref{flow_conservation1}).  In fact, using  \citet[Lemma 1]{bc}, it is possible  to show that $[I-\PP^\top]^{-1}$ is well-defined. Then  $x^*$ is the unique vector  that satisfies $x^*=[I-\PP^\top]^{-1}e_s$ and $f_a=x_i^*\PP(a|A_i^+)$ for all $a\in A_i^+$, $i\neq t.$
\subsection{Path choice and choice aversion} In this section we note that the solution of our recursive choice model can be equivalently written in terms of path choice probabilities.  In doing so, assume that for each path $r\in\R$ the utility associated to it is a random variable defined as
\begin{equation}\label{Path_utility_random1}
\tilde{U}_r=U_r+\epsilon_r\quad\quad \forall r\in \R,
\end{equation}
\noindent where  $U_r=\sum_{a\in r}(u_a-\kappa_{j_a}\log|A_{j_a}^+|)=\sum_{a\in r}u_a-\sum_{a\in r}\kappa_{j_a}\log|A_{j_a}^+|$  and  $\{\epsilon_r\}_{r\in\mathcal{R}}$ is a collection of absolutely continuous random variables satisfying Assumption \ref{Gumbel}.
\smallskip

Under these conditions, the probability of choosing path $r$ is defined as:
\begin{equation}\label{Path_demand_utility1}
\PP_r\triangleq\PP\left(r=\arg\max_{r^\prime\in\mathcal{R}}\{U_{r^\prime}+\epsilon_{r^\prime}\}\right)\quad\forall r\in \R.
\end{equation}

Equations (\ref{Path_utility_random1}) and (\ref{Path_demand_utility1}) jointly define a path choice model over  $\R$, where we again refer to the Gumbel assumption to obtain:
\begin{equation}\label{Logit_Demand1}
\PP_r=\frac{e^{U_r}}{\sum_{r^\prime\in\R}e^{U_{r^\prime}}}\quad\forall r\in \R.
\end{equation}

However, it is well known that the path choice probability $\PP_r$ can be decomposed in terms of the edge probabilities (e.g., \citet{Fosgerauetal2013} and \citet{Gilbertetal2014}). Formally, we have that  for each path $r=(a_1,\ldots,a_K)\in \R$ with $K\geq 2$, the following equality holds
\begin{equation}\label{Markovian_Demand_Logit}
\PP_r=\prod_{k=1}^K\PP(a_k|A_s^+)\PP(a_{k+1}|A_{j_{a_k}}^+)
\end{equation}

The previous characterization will play a key role in later sections.  Intuitively, Eq. (\ref{Markovian_Demand_Logit}) 
establishes that the $\PP_r$ can be decomposed in terms of the recursive choice probabilities. This equivalence allows us to highlight the role and effect  of the terms $\kappa_{j_a}\log|A_{j_a}^+|$ in the path choice probabilities $\PP_r$.
In particular,  Eq. (\ref{Markovian_Demand_Logit}) takes the form
\begin{equation}\label{heteroLogitSize_1}
\PP_r=\frac{e^{u_r-\rho_r}}{\sum_{r^\prime\in\R}e^{u_{r^\prime}-\rho_{r^\prime}}}\quad\quad\mbox{for all $r\in \R$},
\end{equation}
where $u_r=\sum_{a\in r}u_a$ and $\rho_r\triangleq\sum_{a\in r}\kappa_{j_a}\log |A_{j_a}^+|$. 
\subsection{Cycles} \citet{Fosgerauetal2013} discuss the RL when the graph $G$ may contain cycles, a situation where paths may contain loops and be arbitrarily long. They show that, in general, there will be infinitely many potential paths that connect an origin to a destination, but the probability of choosing a particular path is given by the MNL model, where the choice set is discrete but infinite. In particular, under the existence of cycles, the set of paths $\R$ is replaced by the infinite set $\Omega$, which contains all possible paths. Accordingly, the probability of choosing a particular path $r$ is given by:
$$\PP_r=\frac{e^{u_r-\rho_r}}{\sum_{r^\prime\in\Omega}e^{u_{r^\prime}-\rho_{r^\prime}}}\quad\quad\mbox{for all $r\in \Omega$}.$$
From the previous formula, it is easy to see that  we can incorporate cycles in the same way as \citet{Fosgerauetal2013}. However, an important difference with \citet{Fosgerauetal2013} is the fact that $\PP_r$ incorporates  the terms $\rho_r$. In particular, the ratio of choice probabilities between  $r$ and $r^\prime$ depends on the difference $(u_r-\rho_r)-(u_{r^\prime}-\rho_{r^\prime})$.
\section{Choice aversion and the IIA property}\label{s3}

In the context of path choice models, it is well known that the traditional MNL model is restricted by the IIA property, which does not hold in the context of route choice due to the \emph{overlapping paths} problem. The main implication of overlapping paths is that the traditional MNL produces unrealistic path choice probabilities (\citet{BenAkiva_Ramming1998} and \citet{BenAkivaBierlarie1999}).
\smallskip

 In order to solve this problem, the route choice literature  has proposed the Path Size Logit  (PSL) approach. In simple terms, PSL models extend the MNL by adding a  \emph{correction} term to path utilities which account for the degree  of overlapping  among paths. For instance, \citet{BenAkiva_Lerman1985}, \citet{BenAkivaBierlarie1999}, \citet{Frejinger_Bierlarie2003}, and, recently,  \citet{DUNCAN20201} propose different corrections to  the MNL in order to solve the overlapping problem. From a behavioral point of view,  PSL models try to correct the fact that the IIA property should be relaxed in contexts where paths are not distinct or independent.

Following \citet{Luce1959}, the IIA property  can stated as follows: Given $r,r^\prime\in\R$,  
	\begin{equation}\label{IIA}
	{\PP(\mbox{choosing $r$ from $\R$})\over \PP(\mbox{choosing $r^\prime$ from $\R$})}={\PP(\mbox{choosing $r$ from $\tilde{\R}$})\over \PP(\mbox{choosing $r^\prime$ from $\tilde{\R}$})} \quad \mbox{for $\R\subseteq\tilde{\R}$}. 
	\end{equation}

From (\ref{IIA}) it follows that  the IIA property  establishes that the ratio between the probabilities of choosing $r$ and $r^\prime$ is independent of the choice set  containing $r$ and $r^\prime$. In other words, the  comparison between paths $r$ and $r^\prime$ should not be affected by expanding (shrinking) the set of paths, e.g., by adding (severing) links in some nodes.

PSL models described above allow for violations of Eq. (\ref{IIA}). In this section we  show how our recursive model also allows for violations of condition (\ref{IIA}).\footnote{Our model allows for violations of the IIA property at the route level. However, we remark that at the node level, edge choice probabilities satisfy the IIA condition.} Thus, the notion of choice aversion  can be seen as  a  natural mechanism that overcomes the problem of overlapping paths in transportation networks.

 \smallskip
To see how our model works,  for ease of exposition assume $\kappa_i=\kappa$ for all $i\notin \{s,t\}$,  let us rewrite Eq. (\ref{heteroLogitSize_1}) in \S \ref{s2} as follows:
\begin{equation}\label{Logit_size1}
\PP_r=\frac{e^{u_r-\kappa \gamma_r}}{\sum_{r^\prime\in\R}e^{u_{r^\prime}-\kappa\gamma_{r^\prime}}}\quad\quad\mbox{for all $r\in \R$},
\end{equation}
where $u_r=\sum_{a\in r}u_a$, $\gamma_r\triangleq\sum_{a\in r}\log |A_{j_a}^+|$, and $\kappa\geq 0$ is  the (homogeneous)  choice aversion parameter.\footnote{Note that equation (\ref{heteroLogitSize_1}) simplifies to (\ref{Logit_size1}) when the choice aversion parameters are fixed across nodes. In this case, $\rho_r = \kappa \gamma_{r}$.} 
\smallskip

Expression (\ref{Logit_size1}) deserves some comments. First, for $\kappa>0$,  the term $\kappa\gamma_r$ can be seen  as  a penalty term that accounts for the size of choice sets at each of the nodes accessed along path $r$.\footnote{Note that  paths passing through a common  node $i$ would share  the penalty $\kappa\log |A_i^+|$.} However, different from traditional corrections in the PSL literature, $\kappa\gamma_r$ is link additive. Second, 
the penalty  $\kappa\gamma_r$  accounts for the degree of overlapping among different paths. In particular,   adding or severing  links  to  some of the nodes crossed by path $r$ will modify the value of $\kappa\gamma_r$.

To gain some intuition , consider two paths $r$ and $r^\prime$  with associated probabilities $\PP_r$ and $\PP_{r^\prime}$ respectively. Computing the probability ratio between $r$ and $r^\prime$ we get:
\begin{equation}\label{IIA_ratio}
{\PP_r\over \PP_{r^\prime}}= {e^{u_{r}}\over {e^{u_{r^\prime}}}}\times {e^{-\kappa\gamma_r}\over e^{-\kappa\gamma_{r^\prime}}}
\end{equation}

Expression (\ref{IIA_ratio})  shows that the ratio ${\PP_r\over \PP_{r^\prime}}$ depends on the  ratio between the utilities associated to paths $r$ and $r^\prime$ times the ratio between $\kappa\gamma_r$ and $\kappa\gamma_{r^\prime}$.  This latter term  incorporates information about  $r$ and $r^\prime$  regarding  choice sets at each node crossed by these paths. This information is captured by the terms  $\kappa\gamma_r$ and $\kappa\gamma_{r^\prime}$ respectively.

 Concretely, in Eq. (\ref{IIA_ratio}), it is easy to see that  adding or  deleting  links in a particular node crossed by  $r$ (or $r^\prime$)  will affect the ratio ${e^{-\kappa\gamma_r}\over e^{-\kappa\gamma_{r^\prime}}}$, and as a consequence ${\PP_r\over \PP_{r^\prime}}$ will be modified. In other words,  ${\PP_r\over \PP_{r^\prime}}$ is not independent of the set $\R$. Furthermore, we note that when $\kappa=0$,  then ${\PP_r\over \PP_{r^\prime}}= {e^{u_{r}}\over {e^{u_{r^\prime}}}}$ and we recover the IIA property in the MNL model.
\smallskip
\smallskip

It is worth remarking that if paths $r$ and $r^\prime$ cross through the same nodes, we get $\kappa\gamma_r=\kappa\gamma_{r^
\prime}$, so that ${\PP_r\over \PP_{r^\prime}}={e^{u_r}\over e^{u_{r^\prime}}}.$ Thus,  the factor  ${e^{-\kappa\gamma_r}\over e^{-\kappa\gamma_{r^\prime}}}$ captures the degree of overlapping between different paths.
\smallskip

\begin{figure}[h]
\centering
\includegraphics[width=0.49\textwidth]{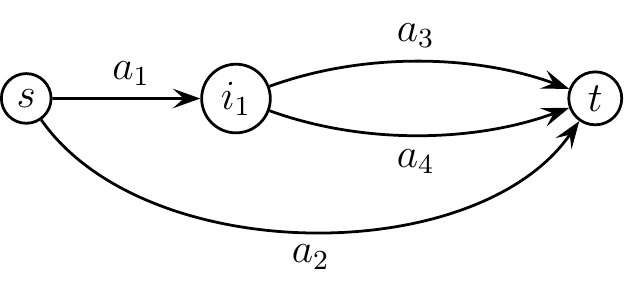}
\caption{Logit path choice.}
\label{fig:Logit_Path_choice}
\end{figure}

In order to see how our model overcomes the overlapping problem,  we  study  a  concrete case.  Let us reintroduce the network in Figure \ref{fig:Logit_Path_choice} where the set of paths is given by $\mathcal{R}=\{r_1,r_2,r_3\}$ with $r_1=(a_1,a_3)$, $r_2=(a_1,a_4)$, and $r_3=(a_2)$. For this example, let $u_a = -c_a$ and assume $c_{a_1}=1.9$, $c_{a_3}=c_{a_4}=0.1$, and $c_{a_2}=2$.%

As we have previously discussed for this network structure,  paths $r_1$ and $r_2$ overlap, sharing the common edge $a_1$.  Under this parameterization of instantaneous utility, coupled with $\kappa=0$, it follows that $u_{r_1}=u_{r_2}=u_{r_3}=-2$, and, consequently, the logit choice rule (\ref{Logit_size1}) assigns one third of flow  to each path. In other words, with $\kappa=0$, we get $\PP_{r_1}=\PP_{r_2}=\PP_{r_3}={1\over 3}$. However, since paths $r_1$ and $r_2$ are identical in utility, the assignment  $\PP_{r_1}=\PP_{r_2}={1\over 4}$ and $\PP_{r_3}={1\over 2}$ is a more appropriate allocation. 
\smallskip

However, as $\kappa \rightarrow 1$, the logit model with choice aversion predicts a flow allocation approaching $\left({1\over 4},{1\over 4}, {1\over 2}\right)$.  To understand this, we  look at the probability ratio between paths $r_1$, 
$r_2$, and $r_3$:
\begin{equation}\label{Ratio_Example}
{\PP_{r_1}\over \PP_{r_2}}=1\quad\text{and}\quad {\PP_{r_1}\over \PP_{r_3}}={\PP_{r_2}\over \PP_{r_3}}=e^{-\kappa\log 2}.
\end{equation}
From the previous expression, it follows that the value of $\kappa$ will affect the ratios ${\PP_{r_1}\over \PP_{r_3}}$ and ${\PP_{r_2}\over \PP_{r_3}}$, but not ${\PP_{r_1}\over \PP_{r_2}}$. In particular, Eq. (\ref{Ratio_Example}) shows that 
the ratios ${\PP_{r_1}\over \PP_{r_3}}$ and ${\PP_{r_2}\over \PP_{r_3}}$ are decreasing in $\kappa$. In other words, as the degree of choice aversion increases, the probabilities $\PP_{r_1}$ and $\PP_{r_2}$ decrease while the probability associated to $r_3$ increases.
\smallskip

\begin{figure}[h]
	\centering
	\includegraphics[width=0.7\linewidth]{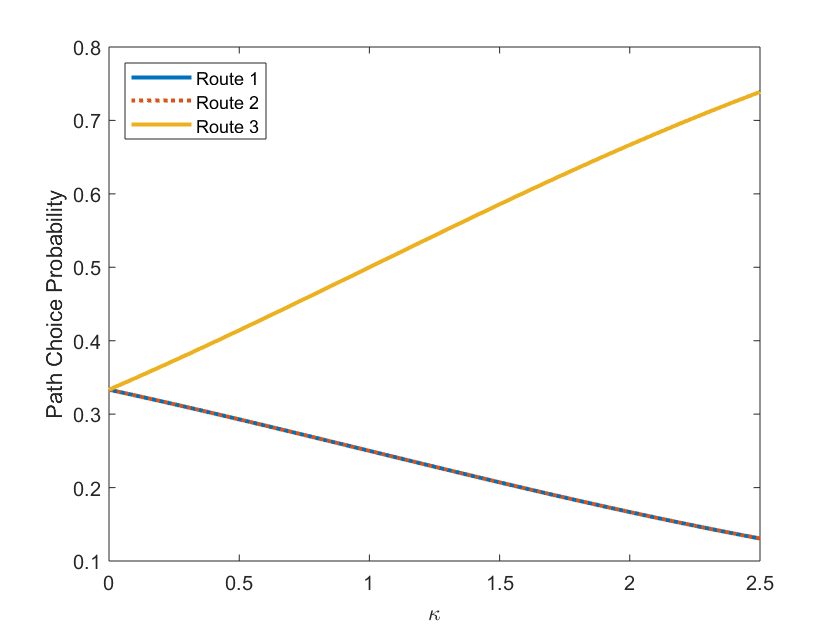}
	\caption{Route Choice Probabilities for Figure \ref{fig:Logit_Path_choice}.}
	\label{fig:figure2example}
\end{figure}

 Figure \ref{fig:figure2example} shows how the route choice probabilities in Figure \ref{fig:Logit_Path_choice} respond to $\kappa \in [0,2.5]$ under choice aversion. 

Now assume that a new edge $\hat{a}$ is added at node $i_1$ which connects to destination node $t$.  This implies that the new set of paths is $\hat{\mathcal{R}}=\mathcal{R}\cup\{\hat{a}\}$. In terms of Eq. (\ref{Ratio_Example}), adding  $\hat{a}$ implies that:
$${\PP_{r_1}\over \PP_{r_2}}=1\quad\text{and}\quad {\PP_{r_1}\over \PP_{r_3}}={\PP_{r_2}\over \PP_{r_3}}=e^{-\kappa\log 3}.$$
This latter expression makes explicit the fact that changes in $\mathcal{R}$ will change the probability ratio between different paths,
i.e., IIA does not hold.
\smallskip

\subsection{Choice aversion compared to the Adaptive PSL model} What the analysis just laid out shows---which applies to the general case of directed networks---is that from the vantage point of path selection, choice aversion is a robust way to derive path choice probabilities, even in the case of overlapping of different routes. This robustness feature makes our approach similar to the class of PSL models, which is widely used in applied work (e.g.  \citet{DUNCAN20201}).
\smallskip

In this section we compare  our approach  with the APSL model, a state-of-the-art extension of the class of PSL models introduced by \cite{DUNCAN20201}.\footnote{For ease of exposition, we provide  the details of the APSL and other PSL models in Appendix \ref{Appendix}.} Figure \ref{fig:figure2fosgerauetal2013} displays a network topology similar to one   featured in \citet{Fosgerauetal2013}. We test the performance of the choice aversion model on this network in calculating path choice probabilities. We assume that this is a directed acyclical graph, where the set of paths is given by $\mathcal{R}=\{r_1,r_2,r_3,r_4\}$ with $r_1=(12,23,35)$, $r_2=(12,23,34,45)$, $r_3=(12,24,45)$, and $r_4=(15)$. Similar to the previous section, let $u_a = -c_a$ for each edge $a$.
\begin{figure}[h]
	\centering
	\includegraphics[width=0.7\linewidth]{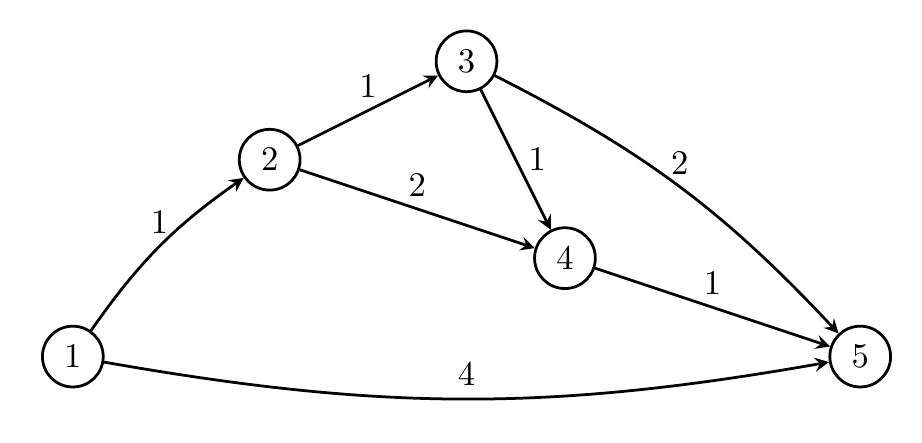}
	\caption{Choice aversion compared to PSL models.}
	\label{fig:figure2fosgerauetal2013}
\end{figure}

This graph represents a more complex uncongested network topology where the cost of all routes $r_i \in \mathcal{R}$ are equal. Thus, the only difference in routes 1 through 4 are the choice sets at each node along the path. The MNL model (equivalent to $\kappa=0$ in the choice aversion model) predicts equal path choice probabilities, i.e., $\PP_{r_i}={1\over 4}$ for $i=1,2,3, 4$. 
\smallskip

\begin{figure}[ht]
	\centering
	\begin{subfigure}[c]{.49\linewidth}
		\centering
		\includegraphics[width=1.1\linewidth]{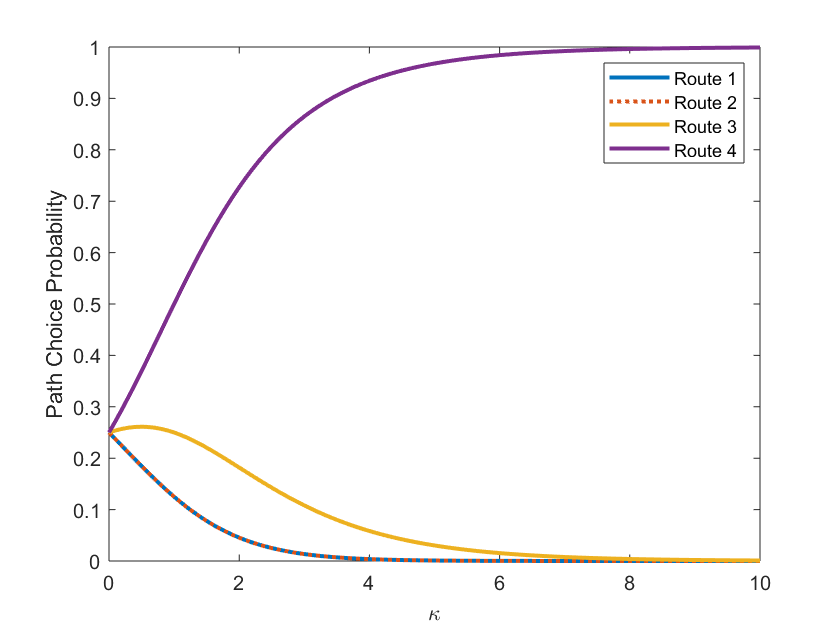}
		\caption{Choice aversion model}
		\label{fig:choiceaversionfosgerau2013}
	\end{subfigure}
	\hspace{0pt}
	\begin{subfigure}[c]{.49\linewidth}
		\centering
		\includegraphics[width=1.1\linewidth]{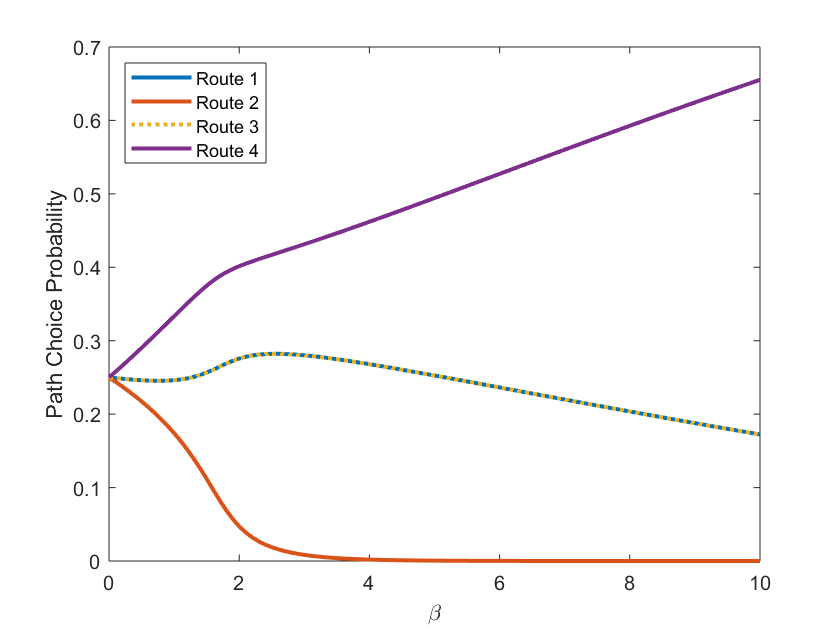}
		\caption{Adaptive PSL model}
		\label{fig:adaptivepslfosgerau2013}
	\end{subfigure}
	\caption{Route choice probabilities for complex network example.}\label{fig:complexnetwork}
\end{figure}

Figure \ref{fig:choiceaversionfosgerau2013} shows choice probabilities as $\kappa$ increases from 0 to 10. As the choice aversion penalization grows larger, $\PP_{r_4}$ approaches 1, since it is the only route with no alternatives for the user to decide between once the user travels beyond node 1. On the other hand, while $r_1$ and $r_2$ have equivalent choice aversion terms, $r_3$ has the advantage of lacking an additional downstream choice in comparison, allowing  $\PP_{r_3} > \PP_{r_1} = \PP_{r_2}$ for $\kappa > 0$ until the choice aversion parameter grows so large that the respective route choice probabilities converge and approach zero.

The prediction in Figure \ref{fig:choiceaversionfosgerau2013}  differs from route choice probabilities generated by the RL in \citet{Fosgerauetal2013} and the APSL model proposed by \citet{DUNCAN20201}. This latter model is shown in Figure \ref{fig:adaptivepslfosgerau2013}. In the Adaptive PSL model, the overlapping path  correction  is based on correlation of routes through link-path incidence rather than a penalization for size of choice sets \emph{along} the path. As a result, we see that for the APSL model, $\PP_{r_1} = \PP_{r_3}$ for all values of $\beta$, not $\PP_{r_1} = \PP_{r_2}$ as in the choice aversion model.
\smallskip

In other words, the APSL model assigns equal probability to paths 1 and 3 because they share equal edge costs and degrees of overlap with other paths, even as the sequence of costs differs. On the other hand, the choice aversion model assigns equal flow to paths 1 and 2 because they are equal in cost and number of outgoing links at path nodes, where we note that one outgoing link results in a choice aversion penalization of zero.\footnote{We thank an anonymous referee by suggesting this paragraph.}
\smallskip
 
We remark that the behavioral nature of the choice aversion model allows for a different \emph{type} of correction than APSL model. The choice aversion and the APSL model work in a similar way in the sense of overcoming the overlapping path problem. However,  the choice aversion model has a simple and clear behavioral interpretation. This feature sets the choice aversion model apart from other APSL and other PSL models both in terms of interpretation and performance.\footnote{We  note that for paths crossing the same nodes, the IIA property holds. To see this, consider paths $r$ and $r^\prime$, crossing exactly the same nodes. Thus in this case $\kappa\gamma_r=\kappa\gamma{r^\prime}$ and  ${\PP_r\over \PP_{r^\prime}}=e^{u_r-u_{r^\prime}}$.   }

\subsection{\citet{Fosgerauetal2013}'s RL  and the overlapping path problem}\label{LS}
The paper by \citet{Fosgerauetal2013} proposes a link additive correction attribute called \emph{Link Size} (LS). The LS correction is similar in spirit to PSL models. However, instead of correcting for  the number of paths sharing a given link, the LS correction considers the expected link flow  as a proxy for the amount of overlap between paths. Our choice aversion correction $\kappa \gamma_r$ is similar to the LS approach in the sense that is link additive, but the mechanism behind the correction of the overlapping paths problem is quite  different. Our correction penalizes choice sets while LS penalizes common links. In other words, we focus at the node level while  \citet{Fosgerauetal2013} focus at the link level.  Thus, both approaches can be seen as alternatives to each other.\\

In addition, we mention that our approach differs from \citet{Fosgerauetal2013} in two important aspects. First, we show that our approach can generate violations of regularity. Second, we show how our model can be useful to characterize changes on welfare when the network is modified. We discuss these two topics in \S \ref{s4} and \ref{s5}, respectively, and we provide an empirical comparison of these models in \S \ref{s6}.
\subsection{\citet{MAI2015100}'s NRL model}\label{s_RNL_IIA} The paper by \citet{MAI2015100} introduces  a Nested  Recursive Logit (NRL) model to overcome the overlapping paths problem. Using an approach similar to the Nested Logit (\citet{McFadden1978}), their model allows for path utilities to be correlated where the links can have different scale parameters. 
\smallskip

 Formally, \citet{MAI2015100} extend the RL model in \citet{Fosgerauetal2013},   by allowing the scale parameter $\mu_a$ of the Gumbel-distributed random variables $\{\epsilon_{a}\}_{a\in A} $ to be link-specific (in contrast with our Assumption \ref{Gumbel}). To see how their approach works,  consider  a  link $k$ and a node $i=j_k$ where $k\in A_i^-$. Let $\mu_{j_k}$ be an link-specific parameter. Accordingly, the continuation value associated to node $j_k$ is defined as:
	\begin{equation}\label{NRL_cost}
	\varphi_{j_k}(V)=\EE\left(\max_{a\in A_{j_k}^+}\{u_a+\varphi_{j_a}+\mu_{j_k}\epsilon_a\}\right)\quad\forall k\in A.
	\end{equation}

	From  (\ref{NRL_cost}) it follows that  the link-specific  scale parameter $\mu_{j_k}$ affects the continuation values. Based on expression (\ref{NRL_cost}), \citet{MAI2015100} show  that  the IIA property can be relaxed using different scale parameters. More importantly, they show how by allowing  correlation between path utilities, the NRL generates  more realistic path choice probabilities.
	\smallskip

Our approach shares the recursive structure of the NRL. However, both models differ in a few key aspects.  First,  the choice aversion term $\kappa_{j_k}\log |A_{j_k}^+|$ modifies the level of the  associated continuation value, while  the link-specific parameter $\mu_{j_k}$ modifies the shape of it. Second, as we discussed above, the inclusion of the choice aversion term allows us to relax the IIA condition.   \citet{MAI2015100} relax the IIA assumption by allowing correlation across stochastic terms $\{\epsilon_a\}$ at each node and through the scale parameters $\{\mu_{j_k}\}$. Thus, choice aversion exploits the topology of the network to relax IIA, while the NRL exploits the correlation between the random variables $\{\epsilon_a\}$ at each node. Third, as we shall see in \S \ref{s4}, our approach allows for the failure of the regularity  property in the path choice probabilities. This pattern is not captured by the NRL model. A final difference is the fact that our approach allows us to characterize how modifications to the network can be welfare-improving or not. In \S\ref{s5}, we discuss this type of exercise.
 \smallskip
 
 It is worth remarking that  \citet{MAI2015100} model the parameter $\mu_{j_k}$ as a function of $|A_{j_k}^+|$. Thus, they are able to capture the effect of $|A_{j_k}^+|$ in the edge and choice probabilities. This latter fact can be seen as an alternative way of capturing the notion of choice aversion. However, there are some key differences between  their approach and our choice aversion model.
 
\begin{figure}[h]
	\centering
	\includegraphics[width=0.49\textwidth]{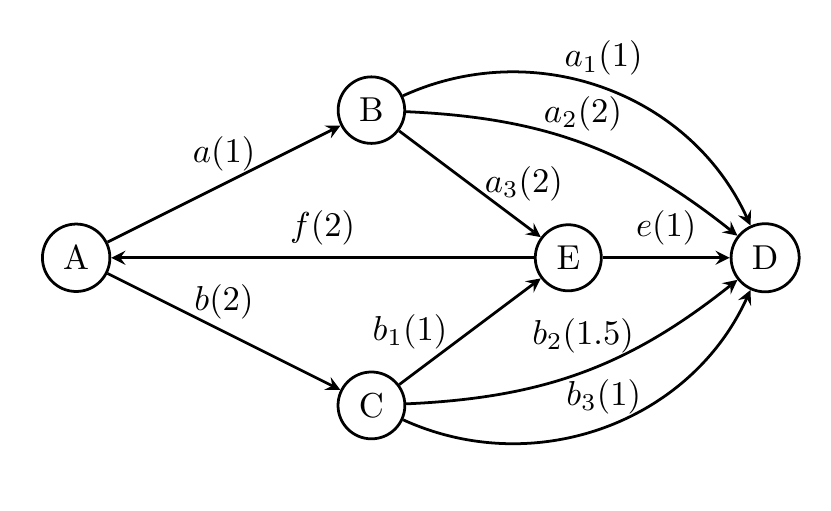}
	\caption{Nested network structure.}
	\label{fig:NRLnested}
\end{figure}

 In order to visualize the differences between the NRL and choice aversion models, we  compare path predictions using the example network shown in Figure \ref{fig:NRLnested}, which was first introduced in \cite{MAI2015100}.

Figure \ref{fig:NRLnested} has five nodes \{A, B, C, D, E\} and 10 links between the source node A and sink node D. There are six paths from A to D: $(a,a_1)$,  $(a,a_2)$,  $(a,a_3,e)$,  $(b,b_1,e)$,  $(b,b_2)$, and $(b,b_3)$. We denote these paths by $r_1$, $r_2$, $r_3$, $r_4$, $r_5$, and $r_6$, respectively. Following \cite{MAI2015100}, link $f$ is not considered in the example routes, but its existence affects the continuation values for overlapping paths in the NRL model. Naturally, in the choice aversion model, link $f$ is included as an option in the choice set at node $E$.

Following \cite{MAI2015100}, we  use the parameterization $\mu_a = 0.5$ and $\mu_b = 0.8$. In addition, we model  $\mu_a$ and $\mu_b$ as follows:
\begin{equation}\label{eq:muform}
\mu_{j_k}= e^{\omega_{j_k} \times {OL_{j_k}}} \; \; \; \mbox{for  $k=a,b,$}
\end{equation}
where $OL_{j_k}\triangleq|A_{j_k}^+|$ refers to the number of outgoing links. Intuitively,  $\omega_{j_k}$ captures how the number of outgoing links affects the magnitude of $\mu_{j_k}$. Given that $\mu_{j_k}\in(0,1]$, we can see from (\ref{eq:muform}) that $\omega_{j_k}=(\log\mu_{j_k}) /OL_{j_k}\leq 0$. \cite{MAI2015100} also provide further empirical evidence that  $\omega_{j_k}<0$. Thus, in order to compare our approach with theirs,  we set $\kappa_{j_k} = -\omega_{j_k}$.
\begin{table}[h]
	\centering
	\resizebox{\columnwidth}{!}{
	\begin{tabular}{ |p{2.5cm}| p{4cm} p{5cm} p{2cm}|  }
		\hline
		\multicolumn{1}{|c|}{Route} & \multicolumn{3}{|c|}{Route Choice Probabilities} \\
		\hline
		& Nested RL  & Choice Aversion & CA\\
		& \ul{($\mu_a = 0.5$, $\mu_b = 0.8$)}  & \ul{($\kappa_B = 0.2310$, $\kappa_C = 0.0744$)} & \ul{($\kappa = 1$)}\\
		$r_1$=($o,a,a_1,d $) & 0.5541 & 0.4227 & 0.4775\\
		$r_2$=($o,a,a_2,d $) & 0.0750 & 0.1555 & 0.1757\\
		$r_3$=($o,a,a_3,d $)& 0.0110 & 0.0572 & 0.0323\\
		$r_4$=($o,b,b_1,d $) & 0.0567 & 0.0679 & 0.0323\\
		$r_5$=($o,b,b_2,d $)& 0.1037 & 0.1120 & 0.1065\\
		$r_6$=($o,b,b_3,d $) & 0.1938 & 0.1847 & 0.1757\\
		\hline
		
	\end{tabular}}
	\caption{Route Choice Probabilities for \cite{MAI2015100}, Figure 4. $\kappa_A$ and $\kappa_B$ are computed using Eq. (\ref{eq:muform}).}
	\label{tab:nrlvsca}
\end{table}

Table \ref{tab:nrlvsca} displays the differences in choice probabilities predicted by NRL and choice aversion models for the network in Figure \ref{fig:NRLnested}. For the choice aversion model, we also include a baseline parameterization where $\kappa=1$ for all nodes $i \neq t$. Here, we see the difference in predictions between the choice aversion model, which penalizes large choice sets directly, and the NRL model, where more options indirectly adjust the scaling of error terms. In other words, we incorporate the effect of a large choice set in a linear way, while \citet{MAI2015100}'s approach is nonlinear.
	
 In terms of path choices,  the NRL predicts a higher choice probability for routes $r_1$ and $r_6$ relative to the choice aversion model. The reason for this comes from the fact that from a behavioral standpoint, when $\omega<0$, the NRL model predicts that the role  of the random shocks in user utilities is smaller. In other words, as the magnitude of $OL_i$ increases,  users' choice of paths becomes \textit{more} precise. This effect is particularly important for paths with a large degree of overlapping. In contrast, the choice aversion model predicts lower likelihoods of selecting the least costly routes as choice sets grow larger, which is aligned with the idea that in larger sets of options, the cost of thinking associated to choice aversion increases (\citet{Ortoleva13} and \citet{FudenbergStra2015}).  

Finally,  we close this section noticing that, given the recursive nature of the choice aversion problem, we can combine our approach with the NRL. To see this note that  Eq. (\ref{NRL_cost}) can be rewritten as:
		\begin{equation*}\label{NRL_cost2}
			\hat{\varphi}_{j_k}(V)=\EE\left(\max_{a\in A_{j_k}^+}\left\{u_a+\hat{\varphi}_{j_a}+\mu_{j_k}\epsilon_a\right\}\right)-\kappa_{j_k}\log|A_{j_k}^+|\quad\forall k\in A.
		\end{equation*}

Studying the properties of combining these two models  is left for future research.
\subsection{\citet{Mai2016} versus choice aversion} \citet{Mai2016} proposes a way to estimate a recursive route choice model, which generalizes  other existing recursive models in the literature. In particular, this approach can incorporate  Multivariate Extreme Value  models at each node, which allows him to accommodate general patterns of substitution between different paths. However, \citet{Mai2016}  does not consider the problem treated in this paper. In particular, his model can generate neither a failure of regularity nor Braess's-like phenomena. Furthermore, and similar to \citet{Fosgerauetal2013} and \citet{MAI2015100}, the model in \citet{Mai2016} predicts that adding links (or paths)  to the network always increases users' welfare.  
\section{Choice aversion and the failure of regularity}\label{s4}
In the standard MNL (i.e., $\kappa_i=0$ for all $i\neq s,t$), adding an additional alternative to the choice set cannot increase the probability that an existing action is selected (and vice versa). This is known as the regularity property in Random Utility models (\citet[Def. 26]{LuceSuppes1965}). 
\smallskip

Formally, in the context of a route choice, the regularity property states that, given the set of routes $\R$ and $\R^\prime$ with $\R\subseteq \R^\prime$, we will have:
	\begin{equation}\label{Regularity} \PP(\mbox{choosing $r$ from $\R$})\geq \PP(\mbox{choosing $r$ from $\R^\prime$})\quad\forall r\in \R\subset \R^\prime.\end{equation}
Intuitively,  condition (\ref{Regularity}) establishes that removing a path from a set $\R^\prime$ should  (weakly) increase the choice probability   of the remaining paths.  Alternatively, adding new paths to the choice set $\R$ should  (weakly) decrease the choice probability of the \emph{existing} paths in $\R$.
\smallskip

\begin{figure}[h]
	\centering
	\includegraphics[width=0.49\textwidth]{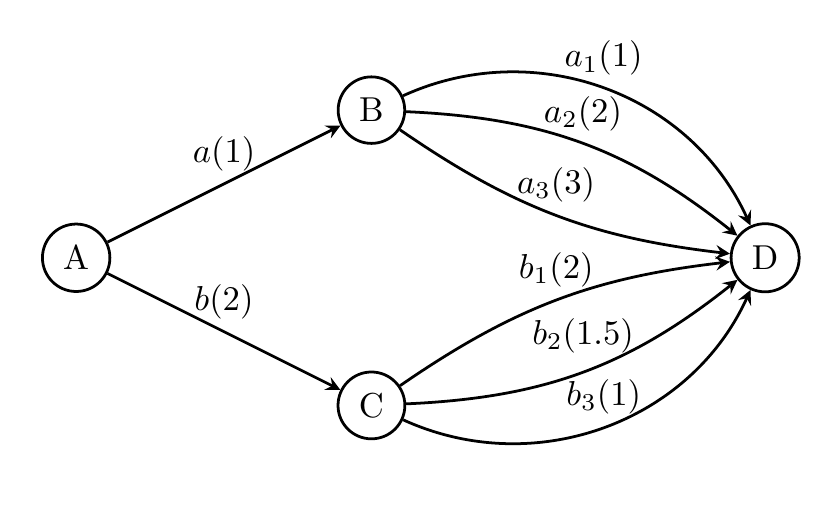}
	\caption{Nested network structure and regularity.}
	\label{fig:nestednetworkstructure}
\end{figure}

  In this section, we show that there exists a critical value of $\kappa_i$, which allows  us to understand how varying the network $G$ can generate  violations of regularity. 
   In order to gain some intuition, we discuss a simple network that allows us to show how regularity may break down. In particular, we study another nested network structure considered  in \citet{MAI2015100}. Figure \ref{fig:nestednetworkstructure} is a simplified version of Figure \ref{fig:NRLnested}, such that links $e$ and $f$ are no longer present. There are still six possible paths from A to D: $(a,a_1)$,  $(a,a_2)$,  $(a,a_3)$,  $(b,b_1)$,  $(b,b_2)$, and $(b,b_3)$. Again, we denote these paths by $r_1$, $r_2$, $r_3$, $r_4$, $r_5$, and $r_6$, respectively.

Tables \ref{tab:Table1} and \ref{tab:Table2} display route choice probabilities when links $a_1$, $a_2$, $b_1$, and $b_2$ are removed from the nested network in Figure \ref{fig:nestednetworkstructure}. Table \ref{tab:Table1} shows the results when choice parameters are homogeneous (i.e., $\kappa_{B}=\kappa_{C}=\kappa=1$), and Table \ref{tab:Table2} shows the results when $\kappa_C=2$, all else held equal. 

\begin{table}[h]
	\centering
	\resizebox{\columnwidth}{!}{
		\begin{tabular}{ |p{2.5cm}| p{1.5cm} | p{3cm} p{3cm} p{3cm} p{3cm} |  }
			\hline
			\multicolumn{1}{|c|}{Route} & \multicolumn{1}{|c|}{Choice Pr.} & \multicolumn{4}{|c|}{Route Choice Probabilities when Edge is Removed} \\
			\hline
			& \ul{$\kappa=1$}  & \ul{$a_1$} & \ul{$a_2$} & \ul{$b_1$} & \ul{$b_2$}\\
			$r_1$ = ($a,a_1$) & 0.4485 & - & 0.6174 {\color{green} (38\%)} & 0.4185 {\color{red} (-7\%)} & 0.4429 {\color{red} (-1\%)} \\
			$r_2$ = ($a,a_2$) & 0.1650 & 0.3726 {\color{green} (126\%)} & - & 0.1539 {\color{red} (-7\%)} & 0.1629 {\color{red} (-1\%)}  \\
			$r_3$ = ($a,a_3$)& 0.0607 & 0.1371 {\color{green} (126\%)} & 0.0836 {\color{green} (38\%)} & 0.0566 {\color{red} (-7\%)} & 0.0599 {\color{red} (-1\%)}  \\
			$r_4$ = ($b,b_1$) & 0.0607 & 0.0914 {\color{green} (51\%)} & 0.0557 {\color{red} (-8\%)} & - & 0.0899 {\color{green} (48\%)}  \\
			$r_5$ = ($b,b_2$)& 0.1001 & 0.1506 {\color{green} (51\%)} & 0.0918 {\color{red} (-8\%)} & 0.1401 {\color{green} (40\%)} & -   \\
			$r_6$ = ($b,b_3$) & 0.1650 & 0.2484 {\color{green} (51\%)} & 0.1514 {\color{red} (-8\%)} & 0.2309 {\color{green} (40\%)} & 0.2444 {\color{green} (48\%)}   \\
			\hline
			
	\end{tabular}}
	\caption{Route Choice Probabilities for Figure \ref{fig:nestednetworkstructure} with $\kappa_{B}=\kappa_C = \kappa = 1$ and Edges Removed.}
	\label{tab:Table1}
\end{table}

\begin{table}[h]
	\centering
	\resizebox{\columnwidth}{!}{
		\begin{tabular}{ |p{2.5cm}| p{1.5cm} | p{3cm} p{3cm} p{3cm} p{3cm} |  }
			\hline
			\multicolumn{1}{|c|}{Route} & \multicolumn{1}{|c|}{Choice Pr.} & \multicolumn{4}{|c|}{Route Choice Probabilities when Edge is Removed} \\
			\hline
			& \ul{$\kappa_{C}=2$} & \ul{$a_1$} & \ul{$a_2$} & \ul{$b_1$} & \ul{$b_2$}\\
			$r_1$ = ($a,a_1$) & 0.5730 & - & 0.7712 {\color{green} (35\%)} & 0.5138 {\color{red} (-10\%)} & 0.5317 {\color{red} (-7\%)} \\
			$r_2$ = ($a,a_2$) & 0.2108 & 0.5535 {\color{green} (163\%)} & - & 0.1890 {\color{red} (-10\%)} & 0.1956 {\color{red} (-7\%)}  \\
			$r_3$ = ($a,a_3$)& 0.0775 & 0.2036 {\color{green} (163\%)} & 0.1044 {\color{green} (35\%)} & 0.0695 {\color{red} (-10\%)} & 0.0720 {\color{red} (-7\%)}  \\
			$r_4$ = ($b,b_1$) & 0.0258 & 0.0453 {\color{green} (75\%)} & 0.0232 {\color{red} (-10\%)} & - & 0.0540 {\color{green} (109\%)}  \\
			$r_5$ = ($b,b_2$)& 0.0426 & 0.0746 {\color{green} (75\%)} & 0.0382 {\color{red} (-10\%)} & 0.0860 {\color{green} (102\%)} & -   \\
			$r_6$ = ($b,b_3$) & 0.0703 & 0.1230 {\color{green} (75\%)} & 0.0630 {\color{red} (-10\%)} & 0.1417 {\color{green} (102\%)} & 0.1467 {\color{green} (109\%)}   \\
			\hline
			
	\end{tabular}}
	\caption{Route Choice Probabilities for Figure \ref{fig:nestednetworkstructure} with $\kappa_C = 2$ and Edges Removed.}
	\label{tab:Table2}
\end{table}
From Table \ref{tab:Table1}, we note that when link  $a_1$ is removed, the choice probabilities of remaining paths \{$r_2$, \dots, $r_6$\} increase. Note that  the probabilities of paths $r_2$ and $r_3$ increase in the same proportion (126\%). Similarly, the probabilities of paths $r_4$, $r_5$, and $r_6$ also increase in the same proportion (51\%). Each increase is even more pronounced in Table \ref{tab:Table2} where $\kappa_{C}=2$. However, in both tables, the increase is not proportional across paths crossing different nodes. This feature comes from the IIA property, which holds within nodes (nests) but not across them. 
\smallskip

Now, consider the case of removing edge $a_2$. In Table \ref{tab:Table1}, the probabilities of paths $r_1$ and $r_3$ increase proportionally (38\%).
However,  the probability of paths $r_4$, $r_5$, and $r_6$ actually \emph{decrease}. This counterintuitive result is a  consequence of the fact that removing edge $a_2$ not only reduces the number of available paths but also decreases the cost associated to paths $r_1$ and $r_3$.  In other words, this effect can be decomposed into two parts.
First, the IIA property implies that the probabilities of $r_{1}$, $r_3$, $r_4$, $r_5$, and $r_6$ will increase  upon removing edge $a_2$. The second force behind this counterintuitive result is that removing edge $a_2$ reduces the choice set when taking paths $r_1$ and $r_3$. For a choice averse user, this latter effect implies that $\kappa\log 3$ reduces to $\kappa\log 2$, which  makes paths $r_1$ and $r_3$ relatively more attractive than  $r_4$, $r_5$, and $r_6$, such that the probabilities of $r_4$, $r_5$, and $r_6$ decrease.  When this second effect dominates, we will observe the failure of regularity associated with removing edge $a_2$.  
\smallskip

It may be tempting to think that the effect in path choices  described above may be driven by the assumption that $\kappa$ is homogeneous. However, Table \ref{tab:Table2} shows that a similar pattern occurs  when we consider $\kappa_B=1$ and $\kappa_C=2$. In this case, the failure of regularity is even more pronounced through relatively larger changes to remaining path probabilities.
\smallskip

We formalize the previous intuition in Proposition \ref{Regularity_Characterization} below. In doing so, recall that  $\R_i$ is the set of paths passing through  node $i$. Similarly, the set of paths not passing through node $i$ is defined as $\R_i^c$. In addition, define $\R_{ia}\subseteq \R$ as  the set of paths passing through node $i$ after removing edge $a$ at node $i$. Finally, let $\PP(\R_i)=\sum_{r\in\R_i}\PP_r$ and $\PP(\R_{ia})=\sum_{r\in \R_{ia}}\PP_r$.
\smallskip
\begin{proposition}\label{Regularity_Characterization}
	The probability of choosing a path $r\in \R_i^c$ decreases after removing an edge $a\in A_i^+$ if
	\begin{equation}\label{Regularity_fail}
	\kappa_i>{\log\left(\PP(\R_{i})\over\PP(\R_{ia})\right)\over \log\left(|A_i^+|\over |A_i^+|-1\right)}.
	\end{equation}
\end{proposition}


Some remarks are in order. First,  Proposition \ref{Regularity_Characterization} provides a simple condition to know when  removing an edge at node $i$ will decrease the probability of paths not crossing node $i$. Condition (\ref{Regularity_fail}) captures the fact that removing an edge at node $i$ will not only modify the set of available paths  but also the  choice aversion cost. Concretely, condition (\ref{Regularity_fail}) establishes a lower bound on the parameter $\kappa_i$ in terms of the path choice probabilities $\PP(\R_i)$ and $\PP(\R_{ia})$  and the magnitude of $|A_i^+|$ and $|A_i^+-1|$. To the best of our knowledge, this result is new to the literature on recursive discrete choice models in transportation networks.

Second, we note that, from a practical point of view,  Proposition \ref{Regularity_Characterization} is useful because it allows us to understand how modifying the structure of the network $G$ can have different effects in terms of flow prediction (choice probability). In particular, (\ref{Regularity_fail}) establishes a simple condition to understand the  choice behavior  as a function of the parameter $\kappa_i$, which is node-specific. Thus, when designing interventions in transportation networks,  Proposition \ref{Regularity_Characterization}  helps us to understand how users' behavior will react to such interventions.
 \smallskip

Third, Proposition \ref{Regularity_Characterization}  predicts that  removing edge $a\in A_i^+$ can decrease the probability of paths passing through nodes different from  $i$.  We have identified this phenomenon as a failure of regularity in the sense of \citet{LuceSuppes1965}. However, behind this result is the factor that  reducing the cardinality of $A_i^+$ reduces the penalization associated with choice aversion. This reduction effect can dominate, making paths crossing node $i$ more attractive than before. As a consequence, the choice probabilities of paths crossing node $i$ should increase while  the choice probability  of paths not crossing node $i$ should \emph{decrease}.

  In the context of rational inattention, \citet{MatejkaMckay2015}  have shown that regularity may fail in the logit model. Our result is different in two aspects. First,  we study a RL model with choice aversion in the context of  transportation networks. Second, our result highlights  the role of  choice aversion by providing a specific condition on $\kappa_i$.   \citet{MatejkaMckay2015} use the idea of information acquisition in order to derive their result.\footnote{We note that \citet{MatejkaMckay2015}'s analysis has been extended to the general class of additive random utility models by \citet{Fosgerauetal2020}.}
\smallskip

Finally, we mention that a particular case of Proposition \ref{Regularity_Characterization} is when $ \kappa_i=\kappa$ for all $i\in A$.
\smallskip

In order to see how Proposition \ref{Regularity_Characterization} applies in the concrete case of the network in Figure \ref{fig:nestednetworkstructure},  Table \ref{tab:TableSummary} summarizes the information after removing edges $a_1$, $a_2$, $b_1$, and $b_2$, respectively. The main message from this table is the simplicity  in checking  Proposition \ref{Regularity_Characterization}.
\begin{table}[h!]
	\begin{center}
		\begin{tabular}{|l|c|} 
			\hline
			Edge Removed & Condition\\
			\hline
			$a_1$ & $\kappa_B > 2.699$\\
			$a_2$ & $\kappa_B > 0.692$\\
			$b_1$ & $\kappa_C > 0.508$\\
			$b_2$ & $\kappa_C > 0.905$\\
			\hline
		\end{tabular} 
	\end{center}
	\caption{Conditions for regularity failure in Figure \ref{fig:nestednetworkstructure}.}\label{tab:TableSummary}
\end{table}

\section{Welfare Analysis and Braess's Paradox}\label{s5}
In  previous sections we have shown how the recursive choice aversion model corrects the problem of predicting routing behavior when there are overlapping paths. Similarly, we have shown how this model may generate violations of regularity when some edge of the network is removed.
\smallskip

 In this section, we show how choice aversion can capture changes to users' welfare when the network topology is modified. Formally, we show that under the presence of  choice aversion, adding edges to the network can decrease users' welfare. In particular, we show how a type of Braess's paradox (\citet{Braess1} and \citet{Braess2}) can emerge even in the case of uncongested networks. 
\subsection{Welfare}\label{s51} Following \citet[Ch. 5]{mcf1}, we define the users' welfare as follows:
\begin{equation}\label{Consumer_Surplus}
\mathcal{W}(\boldsymbol{\kappa})\triangleq\EE\left(\max_{r\in \R}\{U_r+\epsilon_r\}\right)=\log\left(\sum_{r\in \mathcal{R}}e^{U_r}\right),\end{equation}
\noindent where the last equality follows from Assumption \ref{Gumbel} and $\boldsymbol{\kappa}=(\kappa_i)_{i\neq s, t}$. Notice that this definition exploits the equivalence in Eq. (\ref{Markovian_Demand_Logit}) and makes explicit the dependence of users' welfare on the choice aversion parameters $\boldsymbol{\kappa}$.   We say that $\boldsymbol{\kappa}^\prime>\boldsymbol{\kappa}$ iff  $\kappa^\prime_i>\kappa_i$ for all $i\neq s,t.$
\smallskip

Following the literature on discrete choice models, expression (\ref{Consumer_Surplus}) can be interpreted as the inclusive value  of  paths  in $\R$, which is equivalent to say that  $\mathcal{W}(\boldsymbol{\kappa})$ measures  the  inclusive value of the source node $s$.  In particular, $\mathcal{W}(\boldsymbol{\kappa})$ represents the expected utility  faced by the network users.
\smallskip
We now establish some properties of $\mathcal{W}(\boldsymbol{\kappa})$.

\begin{proposition}\label{Properties_W} The function  $\mathcal{W}(\boldsymbol{\kappa})$ satisfies the following properties:
	\begin{itemize}
		\item[a)] 	Let $\boldsymbol{\kappa}^\prime>\boldsymbol{\kappa}>0$. Then $\mathcal{W}(\boldsymbol{\kappa}^\prime)<\mathcal{W}(\boldsymbol{\kappa})$. 
		\item[b)] Consider a node $i\neq t$ and an edge $a\in A_i^+$. Then 
		$${d\mathcal{W}(\boldsymbol{\kappa})\over du_a}=x_i\PP(a|A_i^+)=f_a.$$
		\item[c)] Consider a node $i\neq s,t$.  Then 
		$${d\mathcal{W}(\boldsymbol{\kappa})\over d\kappa_i}=-\sum_{r\in \mathcal{R}_i}\PP_r\log|A_i^+|. $$
	\end{itemize}
\end{proposition}


 Part a) of Prop. \ref{Properties_W} shows an intuitive result: users' welfare decreases when the choice aversion parameter vector $\boldsymbol{\kappa}$ increases. Part b) establishes that an increase in the utility of an edge $a$, $u_a$, subsequently increases users' welfare  $\mathcal{W}(\boldsymbol{\kappa})$ by the amount of flow received by edge $a$. Finally, part c) of Prop. \ref{Properties_W} allows us to predict precisely how much users' welfare $\mathcal{W}(\boldsymbol{\kappa})$ decreases when the choice aversion parameter $\kappa_i$ increases. Quantitatively, $\mathcal{W}(\boldsymbol{\kappa})$ decreases by the sum of the path choice probabilities of all paths passing through node $i$, multiplied by the natural log of the size of the choice set at node $i$, $A_i^+$.
\smallskip

Next, we show how $\mathcal{W}(\boldsymbol{\kappa})$ changes in response to  adding or deleting edges to the network  $G.$ To that end, we connect changes on $\boldsymbol{\kappa}$ with changes on $\mathcal{W}(\boldsymbol{\kappa})$ when the  network $G$ is modified. Formally, we have the following:
\begin{proposition}\label{Braess_prop} Fix a node $i\neq s,t$. Suppose that a new link $a^\prime$ is added to node $i$. Then $\mathcal{W}(\boldsymbol{\kappa})$ increases if and only if the following condition holds:
	\begin{equation}\label{Menu_Condition}
	\PP(a^\prime|A^+_i\cup\{a^\prime\})>1-\left(\frac{|A^+_i|}{|A^+_i|+1}\right)^{\kappa_i}.
	\end{equation}
\end{proposition}



It is worth stressing three implications of this result. First, Proposition \ref{Braess_prop} underscores the way in which local effects\textemdash namely, the addition of alternatives at the node level\textemdash propagate throughout the transportation  network and unlock aggregate welfare effects. Second, it connects the axiomatic characterization of stochastic choice in dynamic settings from \citet{FudenbergStra2015} with transportation networks. In particular, our Proposition \ref{Braess_prop}  extends their Proposition 3 to the case of transportation networks and graph-related settings. Finally, and most notably, it shows that Braess's network paradox may equivalently stem from choice aversion with no allusion to congestion whatsoever.  
\smallskip

From an economic standpoint,  Proposition \ref{Braess_prop} characterizes how adding an edge to the network $G$ is welfare-improving as a function of the value of $\kappa_i$. In particular, after  the new edge $a^\prime$ has been added,  Eq. (\ref{Menu_Condition}) can be easily checked  to know whether   $\mathcal{W}(\boldsymbol{\kappa})$ increases or decreases.
Subsequently, the choice aversion model predicts that there exists a range of values for  the choice aversion parameter $\kappa_i$  where the addition of new edges can lead to a decrease in welfare, even if the utility of newly created routes is higher than existing routes. Thus, we can predict the emergence of Braess's paradox-like phenomena in the case of uncongested transportation networks through Proposition \ref{Braess_prop}. \S \ref{s52} discusses this result in detail.
\smallskip

We note that from an empirical point of view, Eq. (\ref{Menu_Condition}) can be estimated  providing a simple test to understand when modifications  to the network are welfare-improving or not.
\smallskip

Finally, we point out that, in Appendix \ref{appendix:welfarex}, we carry out a welfare analysis comparing our approach with several PSL models, including the APSL. 
\subsection{Braess's paradox as the outcome of choice aversion}\label{s52}  Proposition \ref{Braess_prop} provides a simple condition that characterizes the way in which an improvement at the node level may increase consumer surplus. This result can be connected to Braess's paradox, which states that adding free routes into a transportation network makes every decision maker worse off (\citealp{Braess1,Braess2}).  The connection we establish is that we may generate this phenomenon as a consequence of choice aversion in a general directed network.
\smallskip

\begin{figure}[h!]
	\centering
	\begin{subfigure}[c]{.49\linewidth}
		\centering
		\includegraphics{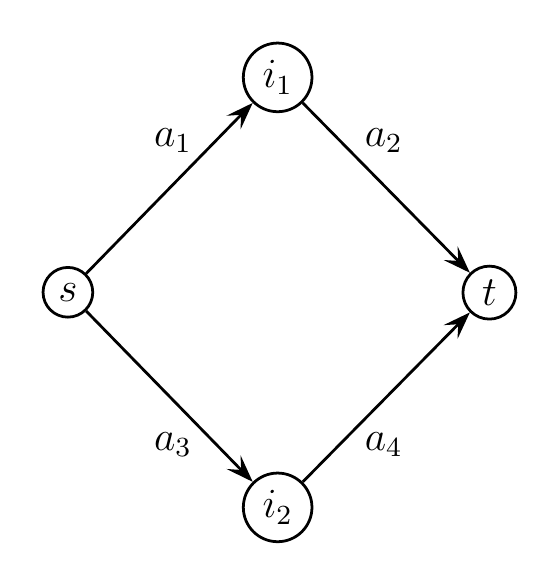}
		\caption{Two unconnected paths}
		\label{fig:Braess-a}
	\end{subfigure}
	\hspace{0pt}
	\begin{subfigure}[c]{.49\linewidth}
		\centering
		\includegraphics{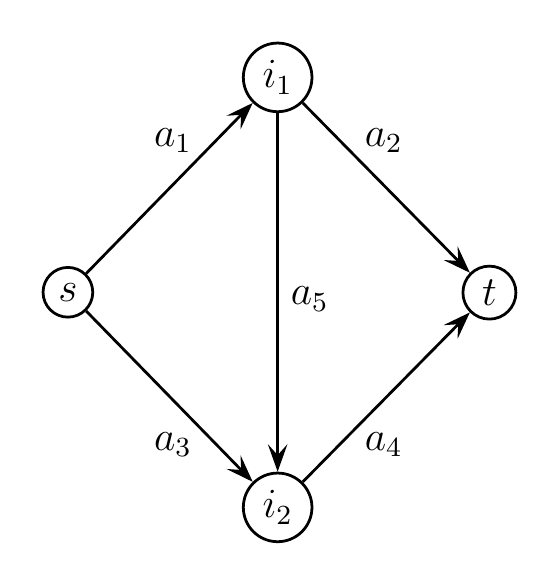}
		\caption{Both paths connected by $a_5$}
		\label{fig:Braess-b}
	\end{subfigure}
	\caption{Braess's paradox ensuing from choice aversion.}\label{Braess}
\end{figure}

	In order to see how our result is related to Braess's paradox,  consider the special directed networks given in Figures \ref{fig:Braess-a} and \ref{fig:Braess-b}. Figure \ref{fig:Braess-a} shows a parallel serial network with paths $r_1=(a_1,a_2)$ and $r_2=(a_3,a_4)$. The set of paths therefore is given by $\R=\{r_1,r_2\}$, and welfare described by
	$$\mathcal{W}(\boldsymbol{\kappa})=\log(e^{U_{r_1}}+e^{U_{r_2}}).$$

Now assume that  the network in Figure \ref{fig:Braess-a} is modified by means of adding a new edge starting at node  $i_1$ and ending at node  $i_2$, as Figure \ref{fig:Braess-b} shows. Now the new set of paths is given by $\tilde{\mathcal{R}}=\{r_1,r_2,r_3\}$, where $r_1$ and $r_2$ are defined as before and $r_3=(a_1,a_5,a_4)$. Users' surplus for this modified network is given by:
$$\tilde{\mathcal{W}}(\boldsymbol{\kappa})=\log(e^{U_{r_1}}+e^{U_{r_2}}+e^{U_{r_3}}).$$
\smallskip
Proposition \ref{Braess_prop} allows us to characterize the effects of adding edge $a_{5}$ into the original network. In particular, $\tilde{\mathcal{W}}(\boldsymbol{\kappa})>\mathcal{W}(\boldsymbol{\kappa})$ iff 
$$\PP(a_{5}|\{a_2,a_{5}\})>1-\left(\frac{1}{2}\right)^{\kappa_{i_1}}.$$
\smallskip
In other words, moving from network \ref{fig:Braess-a} to \ref{fig:Braess-b} improves users' welfare if and only if the probability of choosing $a_5$ is strictly greater than $1-\left(\frac{1}{2}\right)^{\kappa_{i_1}}$, which is automatically satisfied when $\kappa_{i_1}=0$. To state this condition in terms of $\kappa_{i_1}$, we assume  $u=u_{a_1}=u_{a_2}=u_{a_3}=u_{a_4}>0$ and  $u_{a_5}=\epsilon$ with $\epsilon>0$. Given this parameterization, it is easy to see that $\mathcal{W}(\boldsymbol{\kappa})= \log 2e^{2 u}=\log 2+2u$ and 
$\tilde{\mathcal{W}}(\boldsymbol{\kappa})=\log \left(e^{2u}+e^{2u-\kappa_{i_1}\log 2}+e^{2u+\epsilon-\kappa_{i_1}\log 2}\right)=2u+\log\left(1+e^{-\kappa_{i_1}\log 2}(1+e^\epsilon)\right)$.
\smallskip
Then $\tilde{\mathcal{W}}(\boldsymbol{\kappa})>\mathcal{W}(\boldsymbol{\kappa})$ if and only if
$$ {e^{u+\epsilon}\over e^{u}+e^{u+\epsilon}}={e^{\epsilon}\over 1+e^{\epsilon}}>1-\left(\frac{1}{2}\right)^{\kappa_{i_1}}$$ 
which can be rewritten as
$$\kappa_{i_1}<{\log (1+e^\epsilon)\over \log (2)}\triangleq\alpha.$$
It is easy to see that for $\epsilon>0$ we have  $\tilde{\mathcal{W}}(\boldsymbol{\kappa})>\mathcal{W}(\boldsymbol{\kappa})$ if and only if $\kappa_{i_1}<\alpha$.\footnote{Note that for $\epsilon>0$ we have   $\alpha>1$$  $.} 
\smallskip

As we mentioned in \S\ref{s2}, the case of $\kappa_{i_1}\geq 1$ corresponds to a situation where users only want to add alternatives that are sufficiently better that the existing ones. In this case, network users want to add edge $a_5$ whenever $\kappa_{i_1}\in (0,\alpha)$. In particular, for values of $\kappa_{i_1}\geq \alpha$, improving the network can make all users worse off. Thereby, under choice aversion, we can obtain the very same type of paradox described by \citet{Braess1} and \citet{Braess2}. However, while this paradox is usually obtained as a consequence of network users' selfish behavior in a congestion game, we derive it as a consequence of choice aversion, which provides a different perspective to the problem of network design.

\section{Empirical Analysis}\label{s6}

In this section, we move beyond calibration of the choice aversion model and econometrically estimate the choice aversion parameter $\kappa$. Using observed data on paths taken across a given real-world network structure, we can employ maximum likelihood estimation (MLE) procedures to estimate $\kappa$. Previous literature on RL models in transportation networks use MLE routines to analyze the performance and strength of existing models (\cite{Fosgerauetal2013}, \cite{MAI2015100}, \cite{ZIMMERMANN2017183}, \cite{ZIMMERMANN2020100004}). Following this precedence, we exploit a similar approach, incorporating the choice aversion penalization in the utility function of the network user to estimate the value of $\kappa$ alongside other network attributes.\footnote{The MATLAB code used for the empirical exercises in this section are available online or upon request, along with all other code used for this paper.}

Through the following empirical analysis, we demonstrate a set of key strengths of the choice aversion model. First, we show that the obtained estimates for the choice aversion parameter are similar to calibrations given for the simulations on example networks, thus validating the legitimacy of the model and previous inferences from such exercises. Second, we show that the choice aversion model exhibits a similar performance to other RL models in terms of mechanical path prediction on transportation flows. Finally, because of the parsimonious nature of the choice aversion penalty, we show that the model is less computationally demanding, performing at much faster speeds than other corrective additions to the RL model.

\subsection{Illustrative Tutorial  Example in \cite{ZIMMERMANN2017slides} }\label{sec:tut}
\begin{figure}[!h]
	\centering
	\includegraphics[width=0.51\textwidth]{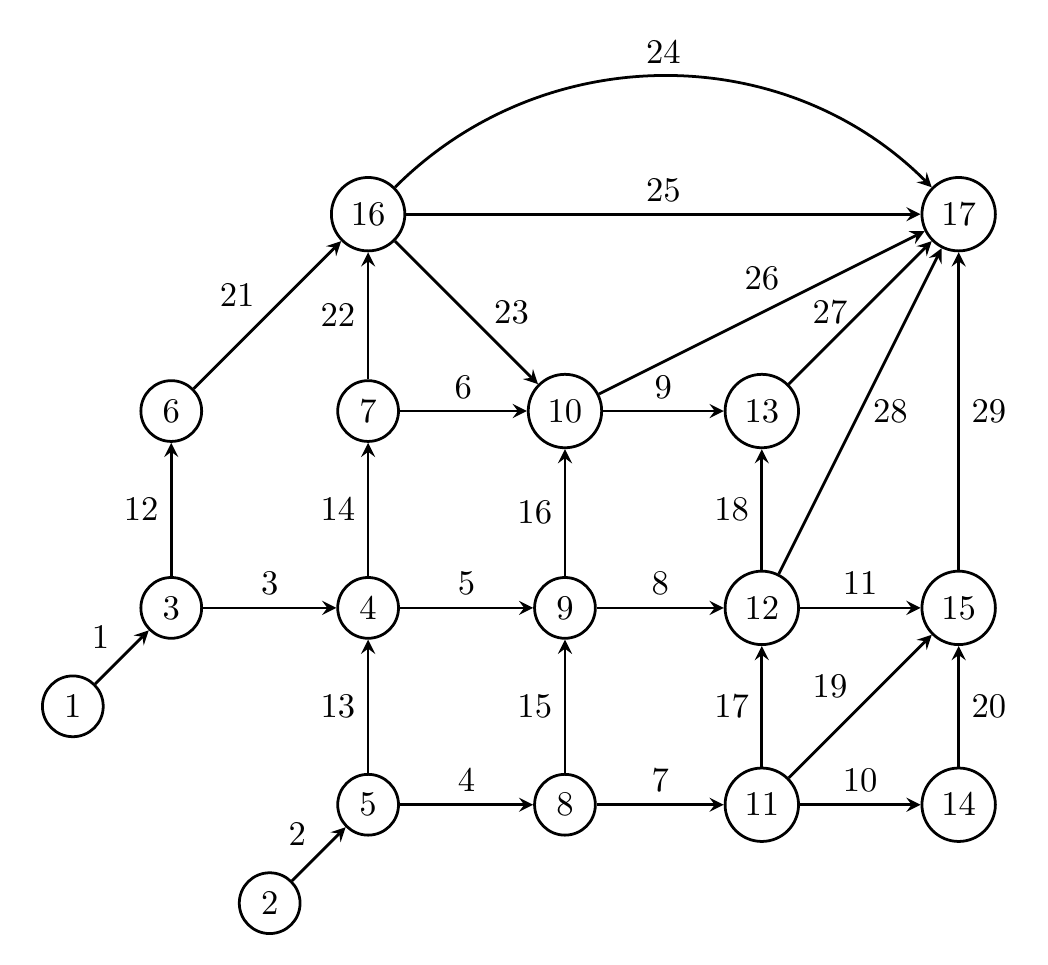}
	\caption{Illustrative Tutorial Network.}
	\label{fig:tutorial}
\end{figure}

We first test the choice aversion model on an illustrative example network used as a tutorial for estimation of the RL model (\cite{ZIMMERMANN2017slides}) and depicted in Figure \ref{fig:tutorial}. This network contains 17 nodes and 29 edges, with 2 origin-destination source-sink pairs (1 to 17 and 2 to 17). There are 100 synthetic path observations over this network generated by a linear utility specification with three attributes: link length, existence of a traffic signal, and a link constant.

For this network, we estimate the traditional RL model, a RL that  incorporates the choice aversion attribute, and a RL model with the Link Size attribute introduced by \cite{Fosgerauetal2013} and discussed in Section \ref{LS}. For the straightforward RL specification, we follow the original authors and estimate coefficients for link length, traffic signal, and a link constant. In addition to this specification, we incorporate the choice aversion term to estimate $\kappa$, the choice aversion parameter.\footnote{For simplicity, we assume that $\kappa$ is homogeneous across nodes in the network. We anticipate future empirical work to relax and test this condition.} For the sake of comparison, we also incorporate the Link Size attribute, which allows users to account for expected flows based on link attributes. We test these specifications both separately and together, as well as allowing for an interaction term between choice aversion and Link Size. These specifications are detailed formally in Eq. (\ref{Tutorial Utility}).\footnote{Note that $u_a + \kappa_a \log |A_{j_a}^+|$ for $\kappa_a < 0$ is equivalent to $u_a - \kappa_a \log |A_{j_a}^+|$ for $\kappa_a > 0$. We have made this modification for the empirical section of the paper to be consistent with testing our hypothesis alongside other attributes in the RL framework.}

\begin{equation}\label{Tutorial Utility}
u_{a}=
\begin{cases}
u_{a,RL}=\beta_{LL} \text{LL}_a + \beta_{TS} \text{TS}_a + \beta_{LC} \text{LC}_a, \\
u_{a,CA}=u_{a,RL} + \kappa_{CA}\log |A_{j_a}^+|, \\
u_{a,LS}=u_{a,RL} + \beta_{LS} \text{LS}_a, \quad\quad\quad\quad\quad\quad\quad\quad\quad\quad\quad\quad\forall a\in A_i^+, \forall i \neq t \\
u_{a,CA\&LS}=u_{a,RL} + \kappa_{CA}\log |A_{j_a}^+| + \beta_{LS} \text{LS}_a, \\
u_{a,CA \times LS}=u_{a,RL} + \kappa_{CA}\log |A_{j_a}^+| + \beta_{LS} \text{LS}_a + \kappa_{CA \times LS}\log |A_{j_a}^+| \times \text{LS}_a
\end{cases}
\end{equation}

Table \ref{tab:tutorial} displays the estimates obtained through the MLE routine conducted across the network for each specification previously mentioned. As may be expected for this tutorial data, the coefficients for link length and traffic signal are the only attributes where significant estimates are consistently observed across specifications.

\begin{table}[!ht] 
	\centering
	\resizebox{\columnwidth}{!}{
		\begin{tabular}{@{\extracolsep{5pt}}lccccc} 
			\\[-1.8ex]\hline 
			\hline \\
			[-1.8ex] & RL & RL-CA & RL-LS & CA \& LS & CA $\times$ LS\\ 
			\hline \\[-1.8ex] 
			$\hat{\beta}_{\text{Link Length}}$ & -2.323 & -2.284 & -2.269 & -2.023 & -2.392\\
			\footnotesize{Std. Error} & (0.304) & (0.301) & (0.475) & (0.362) & (0.564) \\ 
			\footnotesize{t-test} & -7.642 & -7.586 & -5.651 & -5.582 & -4.243 \\ 
			& & & & & \\
			$\hat{\beta}_{\text{Traffic Signal}}$ & -0.337 & -0.253 & -0.297 & 0.235 & -0.275\\
			\footnotesize{Std. Error} & (0.105) & (0.120) & (0.115) & (0.418) & (0.123) \\ 
			\footnotesize{t-test} & -3.225 & -2.103 & -2.575 & 0.562 & -2.232 \\ 
			& & & & & \\
			$\hat{\beta}_{\text{Link Constant}}$ & 0.061 & 0.196 & 0.355 & 0.082 & 0.202 \\
			\footnotesize{Std. Error} & (0.231) & (0.224) & (0.373) & (0.253) & (0.401) \\ 
			\footnotesize{t-test} & 0.264 & 0.875 & 0.951 & 0.325 & 0.505 \\ 
			& & & & & \\
			$\hat{\kappa}_{\text{CA}}$ & --- & -0.539 & --- & -0.559 & -0.261\\
			\footnotesize{Std. Error} & - & (0.364) & - & (0.374) & (0.497) \\ 
			\footnotesize{t-test} & & -1.481 & & -1.495 & -0.525 \\ 
			& & & & & \\
			$\hat{\beta}_{\text{LS}}$ & --- & --- & -0.628 & 0.455 & 0.499 \\
			\footnotesize{Std. Error} & - & - & (0.695) & (0.381) & (1.025) \\ 
			\footnotesize{t-test} & & & -0.904 & 1.194 & 0.487 \\ 
			& & & & & \\
			$\hat{\kappa}_{\text{CA} \times \text{LS}}$ & --- & --- & --- & --- & -1.119\\
			\footnotesize{Std. Error} & - & - & - & - & (1.298) \\ 
			\footnotesize{t-test} & & & & & -0.862 \\ 
			& & & & & \\
			\hline \\[-1.8ex] 
			\footnotesize{N} & 100 & 100 & 100 & 100 & 100 \\ 
			\footnotesize{LL} & 2.374 & 2.361 & 2.370 & 2.354 & 2.357\\
			\footnotesize{Elapsed Time (sec)} & 0.576 & 0.628 & 2.441 & 3.243 & 5.350\\
			\hline 
			\hline \\[-1.8ex] 
	\end{tabular}}
	\caption{MLE Results for Illustrative Tutorial Example}\label{tab:tutorial} 
\end{table} 

In contrast, it does not appear that either choice aversion or Link Size are important attributes to network users under this construction, although they appear to cut into the explanatory power of the preceding variables in some specifications. Given this, we note that the estimated choice aversion parameter $\hat{\kappa}_{CA}$ is consistently negative, while the Link Size attribute ranges between positive and negative values.  Additionally, we note that incorporating the Link Size attribute in the specification dramatically decreases the speed of the MLE routine.

\subsection{Borl{\"a}nge Traffic Network}
Next, we test our model on GPS vehicle data observed in the city of Borl{\"a}nge, Sweden. This traffic data set is well known and has been previously used to test RL models in \cite{Fosgerauetal2013} and \cite{MAI2015100}. Over a span of two years, a sample of 1,832 trips (using a minimum of five links) were observed across a network with 3,077 nodes and 7,459 links (21,452 link pairs). This network includes 466 unique destinations and 1,420 different origin-destination pairs. Most importantly, network users have over 37,000 link choices, making this real-world network a suitable application for testing the choice aversion model. 

Using this data, we proceed with the standard RL model with and without choice aversion and the Link Size attribute as in Section \ref{sec:tut}. The specifications used for the MLE routines in this section are identical to those shown by Eq. (\ref{Tutorial Utility}), except there is no traffic signal attribute to this data.\footnote{\cite{Fosgerauetal2013} and \cite{MAI2015100} include additional network attributes regarding turn angles in their empirical analysis. We omit these from our analysis for the sake of simplicity and clarity.}

\begin{table}[!ht] 
	\centering
	\resizebox{\columnwidth}{!}{
		\begin{tabular}{@{\extracolsep{5pt}}lccccc} 
			\\[-1.8ex]\hline 
			\hline \\
			[-1.8ex] & RL & RL-CA & RL-LS & CA \& LS & CA $\times$ LS\\ 
			\hline \\[-1.8ex] 
			$\hat{\beta}_{\text{Travel Time}}$ & -3.901 & -3.738 & -3.930 & -3.582 & -3.525\\
			\footnotesize{Std. Error} & (-0.127) & (0.124) & (0.136) & (0.145) & (0.141) \\ 
			\footnotesize{t-test} & -30.828 & -30.062 & -28.918 & -24.725 & -25.031 \\ 
			& & & & & \\
			$\hat{\beta}_{\text{Link Constant}}$ & -1.155 & -0.936 & -1.251 & -0.567 & -0.744 \\
			\footnotesize{Std. Error} & (0.015) & (0.049) & (0.018) & (0.040) & (0.056) \\ 
			\footnotesize{t-test} & -77.294 & -18.923 & -69.314 & -14.180 & -13.380 \\ 
			& & & & & \\
			$\hat{\kappa}_{\text{CA}}$ & --- & -0.199 & --- & -0.624 & -0.481\\
			\footnotesize{Std. Error} & - & (0.042) & - & (0.035) & (0.048) \\ 
			\footnotesize{t-test} & & -4.715 & & -17.279 & -10.060 \\ 
			& & & & & \\
			$\hat{\beta}_{\text{LS}}$ & --- & --- & -0.557 & -0.475 & -0.031 \\
			\footnotesize{Std. Error} & - & - & (0.018) & (0.017) & (0.086) \\ 
			\footnotesize{t-test} & & & -31.124 & -27.470 & -0.365 \\ 
			& & & & & \\
			$\hat{\kappa}_{\text{CA} \times \text{LS}}$ & --- & --- & --- & --- & -0.419\\
			\footnotesize{Std. Error} & - & - & - & - & (0.078) \\ 
			\footnotesize{t-test} & & & & & -5.395\\ 
			& & & & & \\
			\hline \\[-1.8ex]
			\footnotesize{N} & 1832 & 1832 & 1832 & 1832 & 1832 \\ 
			\footnotesize{LL} & 5.648 & 5.627 & 4.874 & 4.978 & 4.957 \\
			\footnotesize{Elapsed Time (sec)} & 49.413 & 85.832 & 3738.069 & 5130.600 & 5357.333\\
			\hline 
			\hline \\[-1.8ex] 
	\end{tabular}}
	\caption{MLE Results for Borl{\"a}nge Data}\label{tab:borlange} 
\end{table} 

Table \ref{tab:borlange} displays the results for each specification on the Borl{\"a}nge network. For this data, the estimated choice aversion parameter $\hat{\kappa}_{CA}$ is significant and negative, ranging between $-0.199$ and $-0.628$. This result provides evidence that network users are indeed choice averse and to a degree that is consistent with calibrations presented in Sections \ref{s2}-\ref{s5}. We note that $\hat{\kappa}_{CA}$ and $\hat{\beta}_{LS}$ are similar in range of values and are both significant across specifications, demonstrating that choice aversion has the capability of providing predictive power on transportation networks like Link Size and other additions to the RL model. 

We note that for a complex network such as Borl{\"a}nge, choice aversion does not even double the speed of the MLE routine, while incorporating the Link Size attribute may cause the time required for convergence to increase exponentially. While we also observed a decrease in speed of Link Size for the tutorial data, the decline observed with this data is much sharper moving from the RL to the RL-LS (by over a factor of 75). From an applied perspective, the combination of speed, tractability, and economic interpretation of the choice aversion model should prove attractive when compared to other RL models.

The last column in Table \ref{tab:borlange} displays a coefficient for Link Size that is not significant, while also displaying a negative and significant estimate for the interaction term $\hat{\kappa}_{\text{CA} \times \text{LS}}$. It is intuitive that expected flows and choice aversion are in tandem with one another in terms of affecting individual routing decisions. We interpret this result as an indicator that there is more to understand about the role of choice aversion among users in complex, congested networks. We leave this discussion to be explored more deeply in future empirical and experimental work.

Finally, we note that using the RL-LS rather than the RL-CA
results in a larger improvement of the log-likelihood compared to the RL, suggesting that the RL-LS may outperform the RL-CA in this regard. In Appendix \ref{appendix:oos}, we 
conduct an out-of-sample validation analysis to compare performance across RL models. The main conclusion of this exercise is that the RL-LS and RL-CA exhibit similar out-of-sample performance.

\section{Final Remarks}\label{s7}

The recursive choice model with choice aversion is a highly tractable extension of the standard RL model that can be used to predict reasonable route choice probabilities and provide welfare interpretations in transportation networks. Upon testing our approach  against existing PSL models, our model exhibits the power to provide reasonable corrections and predictions with the benefit of a microfoundation in choice overload. 

In addition, we explore how the choice aversion model allows for a break in the regularity condition typically preserved by path choice probabilities in logit models and extensions. In doing so, we show that, conditional on the degree of choice aversion, removing edges in the network can lead to a decrease in choice probability of certain existing paths.

We also simulate the welfare implications of the choice aversion model and find a novel prediction: even in uncongested networks, a decrease in welfare akin to Braess's paradox can arise when costless edges are added. Here, we also provide a simple characterization for welfare changes conditional on choice aversion which is testable in empirical settings.

Furthermore, we estimate our  model providing evidence that users in transportation networks suffer from choice aversion. An important extension of this work is to consider the role of choice aversion in the context of congested traffic networks and the respective modeling approach. One way to study this question is to extend the model in \citet{bc} by introducing choice aversion in their recursive approach.



Finally, we remark that there is much work to be done regarding the empirical support of choice aversion in transportation networks. We anticipate there is more to understand regarding the role of node-specific (or nest-specific) choice aversion within complex real-world environments. Similarly, we point out that our empirical analysis assumes a homogeneous choice aversion parameter. However, in real-life transportation networks,  it is reasonable to assume that $\kappa$ depends on users' characteristics. We leave for future research the  econometric analysis of this type of model. 

 Future work regarding experiments on behavior of participants in transportation networks would help establish a better understanding of the significance of choice overload in making routing choices. 

\bibliographystyle{apa}
\bibliography{References_Choice_Aversion}

\appendix
\section{Path size logit models}
\label{Appendix}
PSL models include correction terms to penalize routes that share links with other routes, so that the deterministic utility of route $r \in \R$ is $U_r= u_{r}+\mu_{r},$ where $\mu_{r} \leq 0$ is a correction term for route $r \in \R .$ The probability that a user chooses path $r $ is given by:
\begin{equation}\label{Path_Corrections}
\hat{\PP}_{r}=\frac{e^{u_{r}+\mu_{r}}}{\sum_{r^\prime \in R} e^{u_{r^\prime}+\mu_{r^\prime}}}\quad\forall r\in \R
\end{equation}
Following \citet{BenAkivaBierlarie1999},  Path Size Logit (PSL) models adopt the form $\mu_{r}=\beta \ln \left(\gamma_{r}\right),$ where $\beta \geq 0$ is the path size scaling parameter, and $\gamma_{r} \in(0,1]$ is the path size term for route $r \in \R .$ 
\smallskip

Let $u_r =-\theta c_r$, where $c_r = \sum_{a\in r}c_a$ is the instantaneous cost of all edges along path $r$ and $\theta >0$ is a cost scaling parameter. A distinct route with no shared links has a path size term equal to $1,$ resulting in no penalization. Less distinct routes have smaller path size terms and incur greater penalization. The probability that a user chooses route $r \in \R$ is:
\[
\hat{\PP}_{r}=\frac{e^{-\theta c_{r}+\beta \ln \left(\gamma_{r}\right)}}{\sum_{r^\prime \in \R} e^{- \theta c_{r^\prime}+\beta \ln \left(\gamma_{r^\prime}\right)}}=\frac{\left(\gamma_{r}\right)^{\beta} e^{-\theta c_{r}}}{\sum_{r^\prime \in \R}\left(\gamma_{r^\prime}\right)^{\beta} e^{-\theta c_{r^\prime}}}=\frac{1}{\sum_{r^\prime\in R}\left(\frac{\gamma_{r^\prime}}{\gamma_{r}}\right)^{\beta} e^{-\theta\left(c_{r^\prime}-c_{r}\right)}}
\]

The Path Size Logit (PSL) model was first proposed by \citet{BenAkiva_Ramming1998}, and states that the PSL path size term for route $r \in \R, \gamma_{r}^{\mathrm{PS}},$ is defined as follows:
\begin{equation}\label{PSL_term}
\gamma_{r}^{P S}=\sum_{a \in r} \frac{c_{a}}{c_{r}} \frac{1}{\sum_{r^\prime \in \R} \delta_{ar^\prime}}
\end{equation}

where $\delta_{ar^\prime}=1$ if edge $a$ belongs to path $r^\prime$ and $\delta_{ar^\prime}=0$ otherwise.  

In Eq. (\ref{PSL_term}), each link $a$ in route $r$ is penalized (in terms of decreasing the path size term and increasing the cost of the path) according to the number of paths in the choice set that also use that $\operatorname{link}\left(\sum_{r^\prime \in \R} \delta_{ar^\prime}\right),$ and the significance of the penalization is weighted according to how prominent edge $a$ is in route $r$, i.e. the cost of edge $a$ in relation to the total cost of path $r$, $\left(\frac{c_{a}}{c_{r}}\right)$.
\subsection{Generalized Path Size Logit (GPSL)} \citet{BenAkivaBierlarie1999} formulate an alternative PSL model (PSL') that attempts to reduce the contributions of excessively expensive routes to the path size terms of more realistic routes in the choice set. The PSL' model states that the PSL  path size term for route $r \in \R, \gamma_{r}^{PSL^\prime},$ is defined as follows:
\begin{equation}\label{PSLprime_term}
\gamma_{r}^{PSL^\prime}=\sum_{a \in r} \frac{c_{a}}{c_{r}} \frac{1}{\sum_{r^\prime \in \R}\left(\frac{\min \left(c_{r^{\prime\prime}}: r^{\prime\prime} \in \R\right)}{c_{r^\prime}}\right) \delta_{ar^{\prime}}}
\end{equation}
where $\delta_{ar^{\prime}}=1$ if edge is in path $r^\prime$ and $\delta_{ar^\prime}=0$ otherwise.

In Eq. (\ref{PSLprime_term}) the contribution of route $r$ to path size terms is weighted according to the ratio of route $r$ and the cheapest route in the choice set $\left(\frac{\min \left(c_{r^{\prime\prime}}: r^{\prime\prime} \in \R\right)}{c_{r}}\right),$ and hence contributions of high costing routes compared to the cheapest alternative are reduced.

As \citet{Ramming2002} describes, however, when a route is completely distinct its path size term is not always equal to
1 which results in an undesired penalization upon the utility of that route. To combat this, \citet{Ramming2002} proposes the Generalized Path Size Logit (GPSL) model. The GPSL model states that the GPSL path size term for route $r \in \R, \gamma_{r}^{GPSL},$ is defined as follows:
\begin{equation}\label{GPSL}
\gamma_{r}^{GPSL}=\sum_{a \in r} \frac{c_{a}}{c_{r}} \frac{1}{\sum_{r^\prime \in \R}\left(\frac{c_{r}}{c_{r^\prime}}\right)^{\lambda} \delta_{ar^\prime}}
\end{equation}
where $\delta_{ar^\prime}=1$ if  edge $a$ is in path $r^\prime$ and $\delta_{ar^\prime}=0$ otherwise and $\lambda \geq 0$. It is easy to see that  the GPSL model is equivalent to the PSL model when $\lambda=0 .$ In Eq. (\ref{GPSL}) the contribution of route $r^\prime$ to the path size term of route $r$ (the path size contribution factor) is weighted according to the cost ratio between the routes, $\left(\left(\frac{c_{r}}{c_{r^\prime}}\right)^{\lambda}\right),$ and hence the contributions of high costing routes to the path size terms of low costing routes is reduced. $\lambda \geq 0$ is the path size contribution scaling parameter to be estimated.
\subsection{Adaptive Path Size Logit Model (APSL)} In an attempt to improve on existing PSL models and extensions, \citet{DUNCAN20201}  propose an  internally consistent PSL model where all components assess the feasibility of routes according to its relative attractiveness due to travel cost and distinctiveness.  Formally, their correction can be defined as follows:
\begin{definition}\label{Def_APSL}
	The APSL choice probabilities, $\PP^{*},(\text { for a choice set of size } \R)$ are a solution to the fixed-point problem $\PP^{*}=G\left(g\left(\gamma^{APSL}(\PP^*)\right)\right)$ where:

	\begin{eqnarray}
	G_{r}\left(g_{r}\left(\gamma^{APSL}(\PP^*)\right)\right)&=&\tau+(1-N \tau) \cdot g_{r}\left(\gamma^{APSL}(\PP^*)\right)\\
	g_{r}\left(\gamma^{APSL}(\PP^*)\right)&=&\frac{\left(\gamma_{r}^{APSL}(\PP^*)\right)^{\beta} e^{-\theta c_{r}}}{\sum_{r^\prime \in \R}\left(\gamma_{r^\prime}^{APSL}(\PP^*)\right)^{\beta} e^{-\theta c_{r^\prime}}} \\
	\gamma_{r}^{APSL}(\PP^*)&=&\sum_{a \in r} \frac{c_{a}}{c_{r}} \frac{1}{\sum_{r^\prime \in \R}\left(\frac{\PP^*_{r^\prime}}{\PP^*_{r}}\right) \delta_{ar^\prime}} 
	\end{eqnarray}
	\[
	\begin{array}{l}
	
	\forall r\in \R, \quad \forall \PP^* \in D^{(\tau)}, \theta>0, \beta \geq 0,0<\tau \leq \frac{1}{\R}
	\end{array}
	\]
	$D^{(\tau)}=\left\{\PP^* \in \RR_{++}^{\R}: \tau \leq \PP^*_{r} \leq(1-(\R-1) \tau), \quad \forall r \in \R, \sum_{r^\prime\in\R}\PP^*_{r^\prime}=1\right\}$
\end{definition}
Despite the fact that there is no closed-form representation of the choice probabilities for the APSL model, the APSL model corrects for many of the internal consistency issues in route cost and distinctiveness that trouble other PSL models, which makes its recent introduction in the literature particularly useful in working to predict route choice probabilities more appropriately.
\section{Proofs}
\label{appendix:proofs}
\subsection{Proof of Proposition \ref{Regularity_Characterization}}
\label{appendix:prop1}
First, recall that $\R_i$ is the set of paths passing through node $i$, and similarly, the set of paths not passing through node $i$ is defined as $\R_i^c$. We note here that $\R_i,\R_i^c\subseteq\R$  and $\R_i \cup \R_i^c = \R$. Finally, recall that $\R_{ia}\subseteq \R$ is the set of paths passing through node $i$ after removing edge $a$ at node $i$. Note that before and after removing an edge the set $\R_i^c$ is the same.
\smallskip

We remark that after the edge $a$ at node $i$ is removed, the utility of paths in $\R_{ia}$ can be expressed as
$$\bar{U}_r=U_r- \Delta_i\quad\forall r\in\R_{ia}$$
where $\Delta_i\triangleq \kappa_i\left(\log|A_i^+-1|-\log|A_i^+| \right)$. Note that $\Delta_i$ is constant across all paths in $\R_{ia}$. 
Finally, recall that $\PP(\R_i)=\sum_{r\in\R_i}\PP_r$ and $\PP(\R_{ia})=\sum_{r\in \R_{ia}}\PP_r$.
\smallskip

  Without loss of generality, fix a path $r\in \R^c_i$.  Let $\PP_r$  and $\bar{\PP}_r$  be the probability of choosing path $r$ before and after removing edge $a$ at node $i$, respectively.  We want to show under what conditions 
we have $\bar{\PP}_r-\PP_r<0$ for any path $r\in\R^c_i$.\\
\smallskip
Note that $\bar{\PP}_r$ can be written as:
$$\bar{\PP}_r={e^{U_r}\over \sum_{l\in \R_{ia}}e^{U_l-\Delta_i}+\sum_{k\in \R^c_{i}}e^{U_k}}\quad \forall r\in \R_i^c$$
\smallskip
Dividing the numerator and denominator by $\sum_{m\in \R}e^{U_m}$, we find that:
$$\bar{\PP}_r={\PP_r\over \sum_{l\in \R_{ia}}\PP_le^{-\Delta_i}+\sum_{k\in \R^c_{i}}\PP_k}\quad \forall r\in \R_i^c$$
From the previous expression, it follows that  $\bar{\PP}_r-\PP_r$ can be expressed as:
$$\bar{\PP}_r-\PP_r=\PP_r\left({1\over \sum_{l\in \R_{ia}}\PP_le^{-\Delta_i}+\sum_{k\in \R^c_{i}}\PP_k}-1\right)$$
Since $\PP_r >0$, it follows that $\bar{\PP}_r-\PP_r$ is negative iff 
$$\left({1\over \sum_{l\in \R_{ia}}\PP_le^{-\Delta_i}+\sum_{k\in \R^c_{i}}\PP_k}-1\right)<0.$$
Rearranging this expression, we  get 
$$1-\sum_{k\in \R_i^c}\PP_k<\sum_{l\in \R_{ia}}\PP_le^{-\Delta_i}.$$
Using the fact that $\PP(\R_i) = \sum_{l\in \R_i}\PP_l=1-\sum_{k\in \R_i^c}\PP_k$, we get
$${\PP(\R_i)\over\PP(\R_{ia})}<e^{-\Delta_i} .$$
Noting that $e^{-\Delta_i}=e^{\log(|A_i^+|/|A_i^+-1|)^{\kappa_i}}=\left({|A_i^+|\over |A_i^+|-1}\right)^{\kappa_i}$, then we conclude:
$$\kappa_i>{\log\left(\PP(\R_{i})\over\PP(\R_{ia})\right)\over \log\left(|A_i^+|\over |A_i^+|-1\right)}.$$\eproof

The following lemma shows that the utility associated to paths passing through node $i$  can be decomposed in a simple way.
\begin{lemma}\label{Decomposition_Lemma} For each node $i\neq t$ the following holds:
	$$\sum_{r\in\R_i}e^{U_r}=\sum_{r^\prime\in \R_{si}}e^{U_{r^\prime}}e^{\hat{\varphi}_i(V)}$$
	where $\hat{\varphi}_i(V)=\log\left(\sum_{a\in A_i^+}e^{V_a}\right)-\kappa_i\log|A_i^+|$.
\end{lemma}
\proof Let  $r^\prime\in \R_{si}$ be defined as $r^\prime=(a_1,\ldots,a_l)$ with $j_{a_l}=i$, and let $r^{\prime\prime}\in \R_{it}$ be defined as $r^\prime=(a_l,\ldots,t)$. Then it follows that we can construct the utility associated to all paths of the form $r=(r^\prime,r^{\prime\prime})\in \R_i$  as $U_{r^{\prime}}+U_{r^{\prime\prime}}$. In particular,  we have $\sum_{r\in\R_i}e^{U_r}=\sum_{r^\prime\in\R_{si}}\sum_{r^{\prime\prime}\in \R_{it}}e^{U_{r^\prime}+U_{r^{\prime\prime}}}$. Exploiting the recursive structure of the problem, it is easy to see that  $e^{\hat{\varphi}_i(V)}=\sum_{r^{\prime\prime}\in \R_{it}}e^{U_{r^{\prime\prime}}}$.
\smallskip

This latter expression implies that 
$$\sum_{r^{\prime\prime}\in \R_{it}}e^{U_{r^\prime}+U_{r^{\prime\prime}}}=e^{U_{r^\prime}}e^{\hat{\varphi}_i(V)}$$
Adding up over all $r^\prime\in \R_{si}$ we conclude that:
$$\sum_{r\in\R_i}e^{U_r}=\sum_{r^\prime\in \R_{si}}\sum_{r^{\prime\prime}\in \R_{it}}e^{U_{r^\prime}+U_{r^{\prime\prime}}}=\sum_{r\in \R_{si}}e^{U_{r^\prime}}e^{\hat{\varphi}_i(V)}.$$
Previous analysis holds for all node $i\neq t$, then the conclusion follows. \eproof

\subsection{Proof of Proposition \ref{Properties_W}}
\label{appendix:prop2}
a) Consider a path $r\in \R$ and assume that $\boldsymbol{\kappa}^\prime,\boldsymbol{\kappa}>0$. By the definition of $r$, let us set  $U_r(\boldsymbol{\kappa}^\prime)=\sum_{a\in r}(u_a-\kappa^\prime_{j_a}\log|A_{j_a}^+|)$. Then it is easy to see that 
$U_r(\boldsymbol{\kappa}^\prime)<U_r(\boldsymbol{\kappa})$ for all $r\in \R$. In particular $e^{U_r(\boldsymbol{\kappa}^\prime)}<e^{U_r(\boldsymbol{\kappa})}$ and summing over all $r\in \R$ we get:
$$\sum_{r\in\R}e^{U_r(\boldsymbol{\kappa}^\prime)}<\sum_{r\in\R}e^{U_r(\boldsymbol{\kappa})}.$$
Taking the log in the previous expression, we conclude $\mathcal{W}(\boldsymbol{\kappa}^\prime)<\mathcal{W}(\boldsymbol{\kappa})$. \\
b)  By Lemma \ref{Decomposition_Lemma} it is easy to see that $$\mathcal{W}(\boldsymbol{\kappa})=\log\left(\sum_{r\in \R_{i}}e^{U_{r}}+\sum_{r\in \R_i^c}e^{U_{r}}\right)=\log\left(\sum_{r^\prime\in \R_{si}}e^{U_{r^\prime}}e^{\hat{\varphi}_i(V)}+\sum_{r\in \R_i^c}e^{U_r}\right).$$ 
Taking the differential with respect to $du_a$, we get 
$$d\mathcal{W}(\boldsymbol{\kappa})={\sum_{r\in \R_{si}}e^{U_{r^\prime}}e^{\hat{\varphi}_i(V)}\over \sum_{r\in \R}e^{U_r}}{\partial \hat{\varphi}_i(V)\over \partial u_a}du_a.$$
Noting that $x_i={\sum_{r\in \R_{si}}e^{U_{r^\prime}}e^{\hat{\varphi}_i(V)}\over \sum_{r\in \R}e^{U_r}}$ we conclude that:
$${d\mathcal{W}(\boldsymbol{\kappa})\over du_a }=x_i\PP(a|A_i^+)=f_a.$$
c) Using the definitions of $\mathcal{R}_i$ and $\mathcal{R}_i^c$, $\mathcal{W}(\boldsymbol{\kappa})$
can be written as:
$$\mathcal{W}(\boldsymbol{\kappa})=\log \left(\sum_{r\in \mathcal{R}_i}e^{u_r-\rho_r}+\sum_{r^\prime\in \mathcal{R}^c_i}e^{u_{r^\prime}-\rho_{r^\prime}}\right)$$
Computing the differential, we find:
$${d\mathcal{W}(\boldsymbol{\kappa})\over d\kappa_i}=-{1\over \sum_{r\in\mathcal{R}}e^{u_r-\rho_r}}\left(\sum_{r\in \mathcal{R}_i }e^{u_r-\rho_r}\right)\log|A_i^+|.$$
Noting  that ${1\over \sum_{r\in\mathcal{R}}e^{u_r-\rho_r}}\left(\sum_{r\in \mathcal{R}_i }e^{u_r-\rho_r}\right)=\sum_{r\in \mathcal{R}_i}\PP_r$, we conclude that 
$${d\mathcal{W}(\boldsymbol{\kappa})\over d\kappa_i}=-\sum_{r\in \mathcal{R}_i}\PP_r\log|A_i^+|<0.$$

\eproof

\subsection{Proof of Proposition \ref{Braess_prop}}\label{appendix:prop3} The proof of this result exploits the recursive structure of the problem. Let $\tilde{\R}$ denote the set of paths \emph{after} adding a  new link $a^\prime$. Accordingly, define $\tilde{\R}_i$.
Let $\mathcal{W}(\boldsymbol{\kappa})=\log\left(\sum_{r\in\R}e^{U_r}\right)$ and  $\tilde{\mathcal{W}}(\boldsymbol{\kappa})=\log\left(\sum_{r\in\tilde{\R}}e^{U_r}\right)$ be consumer surplus before and after the addition of $a^\prime$, respectively.
Note that $\mathcal{W}(\boldsymbol{\kappa})=\log\left(\sum_{r\in\R_i}e^{U_r}+\sum_{r\in\R^c_i}e^{U_r}\right)$  and $\tilde{\mathcal{W}}(\boldsymbol{\kappa})=\log\left(\sum_{r\in\tilde{\R}_i}e^{U_r}+\sum_{r\in\tilde{\R}^c_i}e^{U_r}\right)$.

Adding edge $a^\prime$ increases consumers' surplus if and only if  the following hold:

$$\log\left(\sum_{r\in\tilde{\R}_i}e^{U_r}+\sum_{r\in\tilde{\R}^c_i}e^{U_r}\right)> \log\left(\sum_{r\in\R_i}e^{U_r}+\sum_{r\in\R^c_i}e^{U_r}\right).$$

Noting that $\tilde{\mathcal{\R}}_i^c=\R_i^c$ previous conditions boils down to:
\begin{equation}\label{comparing_inclusives_values}
\sum_{r\in\tilde{\R}_i}e^{U_r}>\sum_{r\in\R_i}e^{U_r}.
\end{equation}

By Lemma \ref{Decomposition_Lemma} we can write the following:
$$\sum_{r\in \R_i}e^{U_r}=\sum_{r^\prime\in \R_i}e^{U_{r^\prime}+\varphi_i(V)-\kappa_i\log|A_{i}^+|}=\sum_{r\in \R_i}e^{U_r+\varphi_i(V)}|A_{i}^+|^{-\kappa_i}$$
Similarly
$$\sum_{r\in \tilde{\R}_i}e^{U_r}=\sum_{r\in \tilde{\R}_i}e^{U_r+\varphi_i(\tilde{V})-\kappa_i\log(|A_{i}^+|+1)}=\sum_{r\in\tilde{\R}_i}e^{U_r+\varphi_i(\tilde{V)}}(|A_{i}^+|+1)^{-\kappa_i},$$
where $\tilde{V}$ denotes the recursive utility (\ref{recursive_utility}) after adding edge $a^\prime$.

Then Eq. (\ref{comparing_inclusives_values}) can be written as follows:
\begin{equation}\label{comparing_inclusive_values_2}
\sum_{r\in \R_i}e^{U_r+\varphi_i(\tilde{V})}(|A_{i}^+|+1)^{-\kappa}_i>\sum_{r\in \R_i}e^{U_r+\varphi_i(V)}|A_{i}^+|^{-\kappa_i}.
\end{equation}

Previous expression can be written as:
\begin{equation*}
	\left(\sum_{a\in A_i^+}e^{V_a}+e^{V_{a^\prime}}\right)(|A_{i}^+|+1)^{-\kappa_i}>\left(\sum_{a\in A_i^+}e^{V_a}\right)|A_{i}^+|^{-\kappa_i}.
\end{equation*}

Then after some algebra we find:
$$\left({|A_{i}^+|\over |A_{i}^+|+1}\right)^{\kappa_i}>{\sum_{a\in A_i^+}e^{V_a}\over \sum_{a\in A_i^+}e^{V_a}+e^{V_{a^\prime}}}.$$

Using the fact that ${\sum_{a\in A_i^+}e^{V_a}\over \sum_{a\in A_i^+}e^{V_a}+e^{V_{a^\prime}}}=1-\PP(a^\prime|A_i^+\cup\{a^\prime\})$
we can conclude that $\tilde{\mathcal{W}}(\boldsymbol{\kappa})>\mathcal{W}(\boldsymbol{\kappa})$ if and only if 
$$\PP(a^\prime|A_i^+\cup\{a^\prime\})>1-\left({|A_{i}^+|\over |A_{i}^+|+1}\right)^{\kappa_i}.$$
\eproof

\section{Welfare exercises using PSL and the choice aversion model  }
\label{appendix:welfarex}
In this appendix we discuss how, in addition to being able to compute changes on welfare using  the choice aversion model, we also compute changes in welfare using several PSL models.  In particular, for this class of models, the welfare is defined as follows:
\begin{equation}\label{Consumer_Surplus2}
\hat{\mathcal{W}}(\boldsymbol{\theta})\triangleq\EE\left(\min_{r\in \R}\{\hat{U}_r+\epsilon_r\}\right)=\log\left(\sum_{r\in \mathcal{R}}e^{\hat{U}_r}\right),\end{equation}
where $\boldsymbol{\theta}$ is a parameter vector describing the specific PSL model and $\hat{U}_r$ represents the adjusted utility after applying the respective correction. 
\smallskip

It is worth pointing out that the PSL models  are not designed to capture changes on welfare. So,  Eq. (\ref{Consumer_Surplus2}) should  be interpreted as an adapted welfare measure. The reason to consider these measures is to compare how traditional PSL models might be used to compute welfare changes with the choice aversion model.
\subsection{Braess's Paradox}


As we have seen in Section \ref{s52}, the choice aversion model reveals that welfare decreases can arise naturally even when networks are uncongested through increasing choice set cardinality.\footnote{Note that in the case of uncongested networks, users do not get involved in strategic interaction.} In this section of the appendix, we will compare how the choice aversion model handles welfare changes relative to existing PSL models. Let $u_a = -c_a$, and consider the parameterization of Figure \ref{fig:Braess-a}  where $c_{a_1} = c_{a_4} = x$ with $x \in [0,3]$ and $c_{a_2} = c_{a_3} = 1$. As we saw in Section \ref{s5}, this figure depicts the case of a simple parallel serial link network where the set of paths is given by $\mathcal{R}_A=\{r_1,r_2\}$ with $r_1=(a_1,a_2)$ and $r_2=(a_3,a_4)$.

Figure \ref{fig:Braess-b} shows the network with edge $a_5$ added to the directed acyclical graph, connecting $i_1$ to $i_2$. For the purpose of observing Braess's paradox, we set $c_{a_5} = 0$. The set of paths is given by $\mathcal{R}_B = \mathcal{R}_A \cup \{r_3\}$, where $r_3 = (a_1,a_5,a_4)$. We calculate the welfare according to (\ref{Consumer_Surplus}) for \ref{fig:Braess-a} and \ref{fig:Braess-b} and compare the difference.

According to Proposition \ref{Braess_prop}, we can compute the threshold for $\kappa$ that determines when adding links is welfare improving.\footnote{For a concise reading, we refer to $\kappa_{i_1}$ simply as $\kappa$, since it is clear we are referring to node $i_1$ in this example.} In particular, we have that adding a link is welfare-improving when
$$\kappa<{\log\left(1-e^{-x}/(e^{-1}+e^{-x})\right)\over \log(1/2)}.$$
After some algebra, we can express a relationship between $\kappa$ and $x$ as follows:
\begin{equation}\label{kappa_x_relationship}
\kappa<-\log\left({e^{x-1}\over e^{x-1}+1}\right)/\log 2.
\end{equation}
From (\ref{kappa_x_relationship}) it is easy to see that when $x=0$ adding an edge is welfare improving when $\kappa<1.89 $, approximately. Similarly, when $x=1$,
the welfare increases when $\kappa<1$. In particular, Eq. (\ref{kappa_x_relationship}) shows that there is an inverse relationship between  $\kappa$ and $x$.

\begin{figure}[h]
	\centering
	\includegraphics[width=0.7\linewidth]{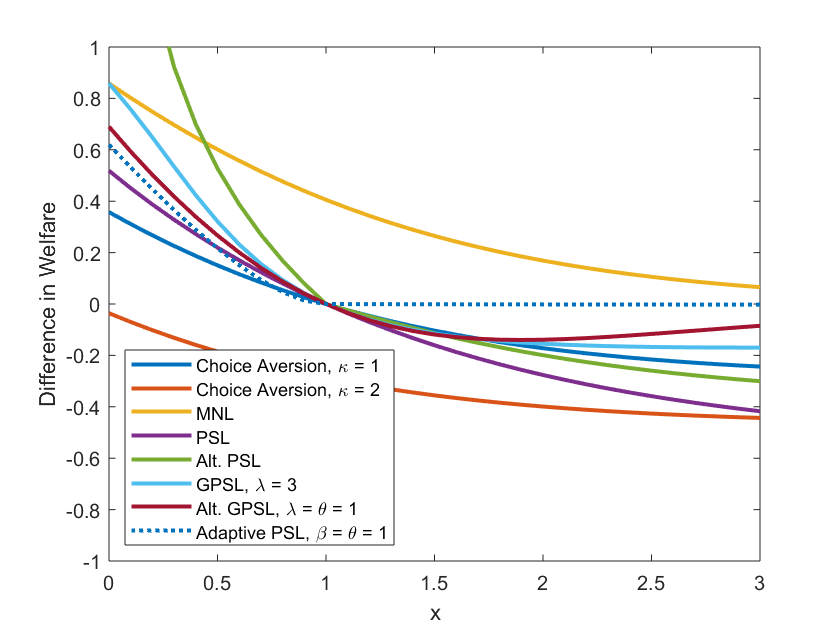}
	\caption{Welfare change from network Figure \ref{fig:Braess-a} to \ref{fig:Braess-b} for various logit models.}
	\label{fig:welfarediffwrtx}
\end{figure}

Based on this observation, we show how $\mathcal{W}(\boldsymbol{\kappa})$ changes when $x$ varies for a constant $\kappa$. Figure \ref{fig:welfarediffwrtx} shows the welfare change for the choice aversion model with $\kappa=\{1,2\}$ 
as well as the MNL model and other PSL models and extensions mentioned in Section 4. 
\smallskip

The choice aversion model where $\kappa=1$ matches this characterization, reflecting a welfare upgrade for $x<1$ and a welfare downgrade when $x>1$. 

In contrast, the MNL model only reflects a nonnegative change in welfare from adding edge $a_5$, a symptom of a model where additional gains in welfare are realized from any additional route added to the choice set of the agent. From a theoretical standpoint, this feature of the MNL model captures the \emph{preference for flexibility} in users' preferences (\citet{Kreps1979}).

Interestingly, most PSL models and extensions shown in Figure \ref{fig:welfarediffwrtx} reflect a welfare change similar to each other: a initial welfare gain which decreases in $x$, ultimately leading to a decrease in welfare. 

However, \citet{DUNCAN20201}'s Adaptive PSL model breaks from the welfare change pattern observed by other models. For this model, the difference in welfare is positive and decreasing until $x=1$, but for $x>1$, the change in welfare is zero. This is a compelling result which speaks to the strengths of this PSL extension.\footnote{To understand why the APSL  differs from traditional PSL models, we note that $c_{r_3} \leq c_{r_1} = c_{r_2}$ when $x \leq 1$, with the equality holding only for $x=1$. Here, it is reasonable that we would observe an increase in welfare when $a_5$ is added to the network: for an uncongested network, we are adding a cheaper route choice for users, and it is intuitive that this would improve welfare. Indeed, this result is consistent with most models as shown in Figure \ref{fig:welfarediffwrtx}. However, for $x>1$, where $c_{r_3} > c_{r_1} = c_{r_2}$, the APSL model reflects no decrease in welfare. Again, this is intuitive: when $x>1$, adding costless edge $a_5$ does not provide users with a cheaper route choice. Since the users would incur cheaper costs from choosing routes $r_1$ or $r_2$, the APSL model considers $r_3$ to be an irrelevant addition to the choice set and thus would not contribute to, nor detract from, welfare.}
\smallskip

As we have discussed in great detail already, an important property of the choice aversion model is its microfoundation based in the behavioral concept of choice overload, which provides justification for the penalization term in the utility function as well as the observed outcomes. While there is a similar welfare difference observed among the choice aversion model with $\kappa=1$ and other PSL models and extensions, this microfoundation sets the choice aversion model apart from other models.

It is important to note the contrast in welfare change between $\kappa=1$ and $\kappa=2$ for the choice aversion model. For $\kappa=2$, Figure \ref{fig:welfarediffwrtx} clearly shows that there is no value of $x>0$ such that a positive welfare change will be observed. Again, the decrease in welfare stems from a dominating penalization on path utilities by the choice aversion term, amplified by the choice aversion parameter $\kappa$. This contrast is important and draws our attention to how welfare changes as $\kappa$ varies for a fixed $x$.

\begin{figure}[h]
	\centering
	\begin{subfigure}[c]{.49\linewidth}
		\centering
		\includegraphics[width=1.1\linewidth]{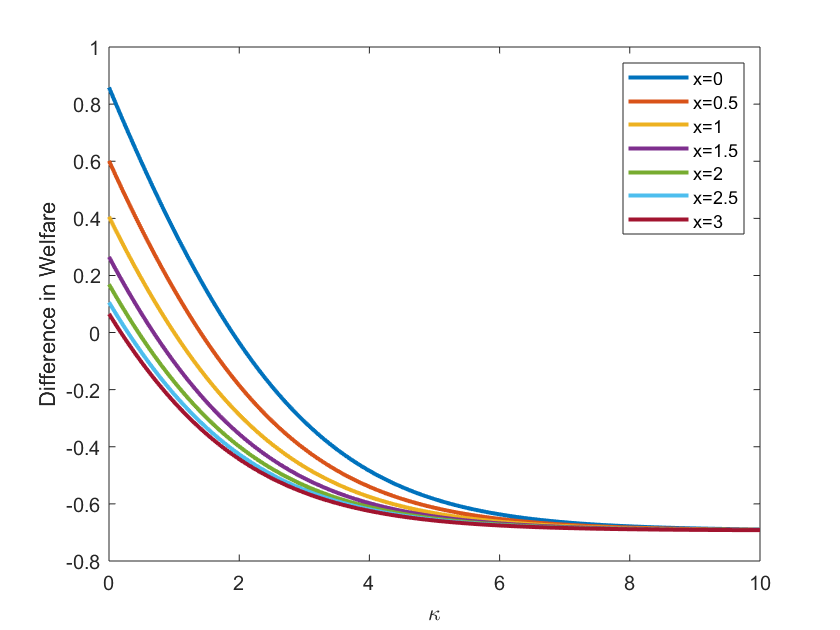}
		\caption{Choice aversion model}
		\label{fig:welfarediffwrtkappa}
	\end{subfigure}
	\hspace{0pt}
	\begin{subfigure}[c]{.49\linewidth}
		\centering
		\includegraphics[width=1.1\linewidth]{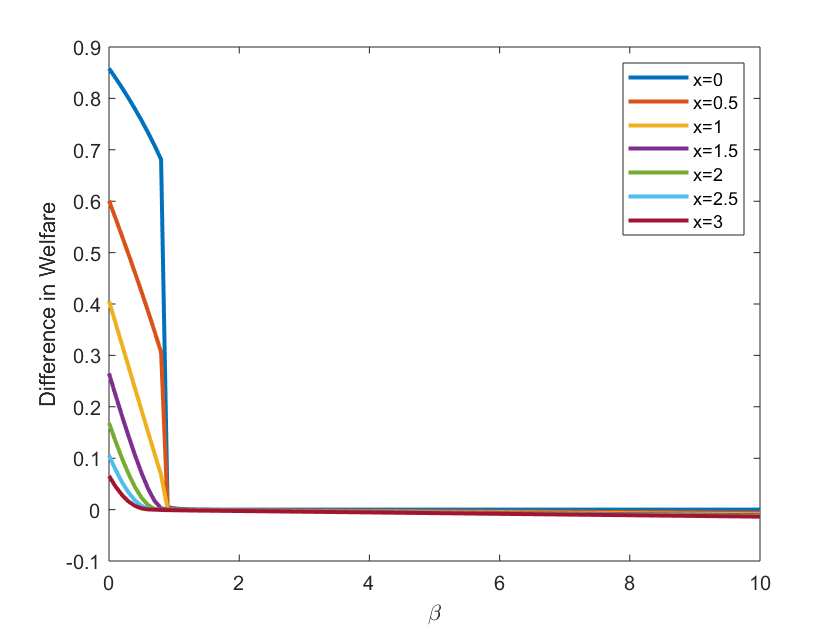}
		\caption{Adaptive PSL model}
		\label{fig:adaptivewelfarediffwrtbeta}
	\end{subfigure}
	\caption{Welfare change as $\kappa$ ($\beta$) varies.}\label{fig:welfarediff}
\end{figure}

Figure \ref{fig:welfarediffwrtkappa} shows how the difference in welfare calculated by the choice aversion model responds to an increase in $\kappa$ from 0 to 10 for fixed $x = \{0,0.5,1,1.5,2,2.5,3\}$. For each value of $x$ displayed, we first see a positive difference in welfare for low values of $\kappa$. As $\kappa$ increases, the difference in welfare continues to decrease until the welfare change is negative. In other words, we see that the threshold $\kappa$ (i.e., the $\kappa$ where the welfare change from adding edge $a_5$ is no longer positive) decreases in $x$. This occurrence stems from the fact that for low values of $\kappa$, the instantaneous cost of each route dominates the choice aversion term in the users' utility functions. Thus, for low values of $x$, the welfare change is initially positive, but as $\kappa$ increases, the choice aversion term is updated with increasing weight until the user's aversion to choice dominates any increase in utility from an additional route created by $a_5$. 
\smallskip

While the choice aversion model predicts this welfare decrease as a result of its behavioral motivation, other logit models may not provide the same outcome. For example, the APSL model does not show a welfare decrease for any value of $x$ as shown by Figure \ref{fig:adaptivewelfarediffwrtbeta}. Rather, this model seems to predict that the welfare change is nonnegative for all values of $\beta \in [0,10]$ and only positive for $\beta <1$.\footnote{Despite the difference in  behavioral foundation, $\kappa$ in the choice aversion model and $\beta$ in most PSL models, including the APSL model, serve a similar purpose of correcting  the degree of overlapping between routes sharing common links. However, quantitatively we cannot compare $\kappa$ and $\beta$.} While this may be sensible for $x>1$, we would expect that a model looking to explain decision-making behavior accurately would display a continuous tradeoff between route costs and aversion to increasing choice set cardinality. However, we see that the threshold $\beta$, where the welfare change is no longer positive, is approximately 1 for all values of $x$ featured and most notably for $x<1$. The APSL model does not seem to be capable of taking into account the tradeoff mentioned above, perhaps as a result of its corrective nature and intentional design. While the APSL model and other PSL models may have a strength in producing more intuitive route choice probabilities, there appears to be a weakness in predicting reasonable welfare changes from a behavioral point of view.

To summarize, while PSL models and extensions are developed to correct the outcome of a standard logit model and provide what may be deemed as more reasonable choice probabilities for a given network topology, what they often lack is the behavioral foundation for any penalties or adjustments in the cost function of the network user, which would justify the outcomes they provide. Their designs also limit their capability to model changes in welfare with respect to internal parameters. This is not to say that the existing PSL models are inferior; indeed, these models are powerful and useful in various applications at providing reasonable predictions in network flow allocations.\footnote{For instance, \citet{DUNCAN20201} analyze an example similar to Figure \ref{fig:Logit_Path_choice}. Their conclusions, while being quantitatively  different,  agree with the predictions made by the choice aversion model.} We simply wish to speak to the strengths of the choice aversion model in the context of applications in transportation networks where choice overload is a factor in users' decision-making, as well as encourage a convergence of the PSL literature with choice aversion models in transportation network applications.

\section{Out-of-Sample Validation}
\label{appendix:oos}

Accompanying our empirical results, we also perform out-of-sample validation analysis for the models compared in Section \ref{s6}. We follow a similar procedure to the existing RL literature (e.g., \cite{MAI2015100}): we generate 40 unique random samples from the GPS vehicle data observed in the city of Borl{\"a}nge, Sweden. Half of the random samples include 80\% of the total path observations and are used to train the models and estimate parameters using the MLE routine from Section \ref{s6}. The remaining 20 testing samples contain 20\% of the total path observations and are held out to calculate predicted log-likelihood using the estimated parameters from the training samples.

Following \cite{MAI2015100}, we use the predicted log-likelihood to compute the test errors

\begin{equation}
	\text{TestError}_i = - \frac{1}{|TS_i|} \sum_{r \in TS_i} \log \PP(r, \hat{\beta}_i)
\end{equation}

where $|TS_i|$ is the size of testing sample $1 \geq i \geq 20$ and $\hat{\beta}_i$ is the set of estimated coefficients from the training sample corresponding with $i$. Put simply, we weight the negative predicted log-likelihood by the size of the testing sample to calculate the test error.

\begin{figure}[h]
	\centering
	\includegraphics[width=0.7\linewidth]{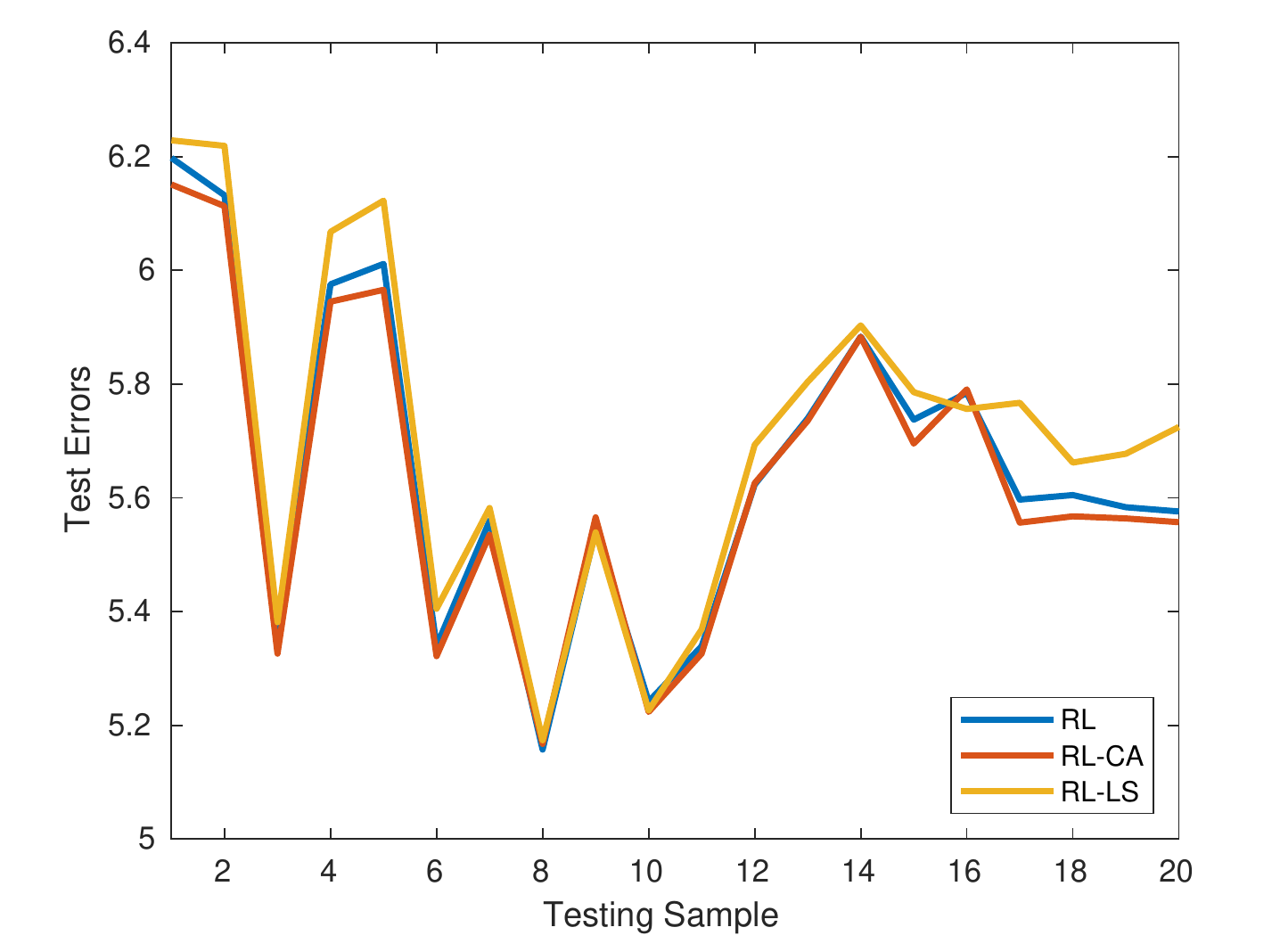}
	\caption{Test Errors of Predicted Log-Likelihood Values for Holdout Samples}
	\label{fig:testerrors}
\end{figure}

Figure \ref{fig:testerrors} displays the test errors for each testing sample when estimating the training sample with each model: the standard RL with no corrections, the RL with choice aversion penalization, and the RL with Link Size (\cite{Fosgerauetal2013}). We note that the RL-CA demonstrates a similar performance to the RL and RL-LS models, with either a test error equal to or below RL and RL-LS errors for most holdout samples.

\end{document}